\documentclass[vecphys]{svmult}

\usepackage{makeidx}         

\usepackage{epsfig}

\usepackage{multicol}        

\usepackage[bottom]{footmisc}

\def\mapbelow#1{\smash{\mathop{\longrightarrow}\limits_{#1}}}

\newcommand{\lab}{\label}

\newcommand{\bc}{\begin{center}}
\newcommand{\ec}{\end{center}}
\newcommand{\be}{\begin{equation}}
\newcommand{\ee}{\end{equation}}
\newcommand{\bea}{\begin{eqnarray}}
\newcommand{\eea}{\end{eqnarray}}
\newcommand{\bs}{\begin{subequations}}
\newcommand{\es}{\end{subequations}}
\newcommand{\beq}{\begin{eqalignno}}
\newcommand{\eeq}{\end{eqalignno}}

\newcommand{\half}{\frac{1}{2}}

\def\lab{\label}
\def\lan{\langle}

\def\pa{\partial}
\def\ran{\rangle}

\def\ri{\right}
\def\ti{\tilde}

\def\le{\left}
\def\al{\alpha}

\def\ga{\gamma}
\def\Ga{\Gamma}
\def\de{\delta}

\def\ka{\kappa}

\def\om{\omega}
\def\Om{\Omega}
\makeindex             

\begin{document}

\title*{Links.\\
Relating different physical systems through the common QFT
algebraic structure}
\titlerunning{Links}
\author{Giuseppe Vitiello
}
\institute{Dipartimento di Fisica "E.R.Caianiello" and INFN\\
Universit\'a di Salerno, 84100 Salerno, Italia\\
\texttt{vitiello@sa.infn.it}
}
%
%
\maketitle

\abstract{In this report I review some aspects of the algebraic
structure of QFT related with the doubling of the degrees of
freedom of the system under study. I  show how such a doubling is
related to the characterizing feature of QFT consisting in the
existence of infinitely many unitarily inequivalent
representations of the canonical (anti-)commutation relations and
how this is described by the $q$-deformed Hopf algebra. I consider
several examples, such as the damped harmonic oscillator, the
quantum Brownian motion, thermal field theories, squeezed states,
classical-to-quantum relation, and show the analogies, or links,
among them arising from the common algebraic structure of the
$q$-deformed Hopf algebra.}

\section{Introduction}
\label{sec:1}


Since several years I am pursuing the study of the vacuum
structure in quantum field theory (QFT) through a number of
physical problems such as boson condensation and the infrared
effects in spontaneously broken symmetry gauge theories, coherent
domain formation and defect formation, soliton solutions, particle
mixing and oscillation, the canonical formalism for quantum
dissipation and unstable states, the quantization in curved
background, thermal field theories, quantum-to-classical
relationship. In this paper I would like to share with the reader
the satisfying feeling of a unified view of several distinct
physical phenomena emerging from such a study of the QFT vacuum
structure. Besides such a pleasant feeling, there is a concrete
interest in pointing out the analogies ("links") among these
phenomena, which arises since these links provide a great help not
only in the formulation of their mathematical description, but
also in the understanding of the physics involved in them. Such a
'compared study' also reflects back to a deeper understanding of
structural aspects of the same QFT formalism.

Quite often QFT is presented as an extension of quantum mechanics
(QM) to the relativistic domain. Sometimes it is referred to as
"second quantization". Of course, the reasons for that come from
the historical developments in the formulation of the quantum
theory of elementary particle physics and solid state physics.
However, a closer view to the formalism of QFT shows that it is
not necessarily related with the relativistic domain and it is not
simply a "second" quantization recipe subsequent the quantization
procedure in QM. For example, the QFT formalism is widely used,
with great success, in condensed matter physics, e.g. in the
formulation of superconductivity, of ferromagnetism, etc., where
typically one does not refer to the relativistic domain. On the
other hand, in dealing with fermion fields one cannot rely on the
quantization scheme adopted in QM for boson creation and
annihilation operators.

As it will appear in the following, QFT is drastically different
from QM. The main reason for this resides in the fact that the
well known von Neumann theorem, which characterizes in a crucial
way the structure of QM \cite{vonneumann, UVQM}, does not hold in
QFT. In QM the von Neumann theorem states that for systems with a
finite number of degrees of freedom all the representations of the
canonical commutation relations (ccr) are unitarily equivalent.
This means that they are physically equivalent; namely, the
representations of the ccr are related by unitary operators and,
as well known, physical observables are invariant under the action
of unitary operators. Their value is therefore the same
independently of the representation one choses to work in. Such a
choice is thus completely arbitrary and does not affect the
physics one is going to describe. The situation is quite different
in QFT where the von Neumann theorem does not hold. Indeed, the
hypothesis of finite number of degrees of freedom on which the
theorem rests is not satisfied since fields involve by definition
infinitely many degrees of freedom. As a consequence, infinitely
many unitarily inequivalent (ui) representations  of the ccr are
allowed to exist \cite{Bratteli,Sew,Umezawa}. The existence of ui
representations is thus a characterizing feature of QFT and a full
series of physically relevant consequences follows.

One of the aspects I will discuss below is related with the
algebraic structure of QFT. I will show that the relevant algebra
underlying the QFT formalism is the Hopf algebra, and this
underlies the existence of the ui representations. It manifests in
the doubling of the system degrees of freedom and its
q-deformation bears deep physical meaning. In the first part of
the paper, I will start by considering some aspects of the
two-slit experiment. This is a typical subject in QM where quantum
features fully show up. The discussion turns out to be useful for
the subsequent discussion of the $q$--deformed Hopf algebra
structure of QFT \cite{25.,Celeghini:1991km} and it also provides
a good example where the quantum-to-classical relation manifests
itself.

The $q$-deformation of the Hopf algebra will be shown to be also
related with quantum dissipation and  with thermal field theory,
where the description of statistical thermal averages of
observables in operatorial terms is made possible by exploiting
the existence of infinitely many ui representations \cite{Umezawa,
TFD, Celeghini:1998sy, Vitiello:2003me}. Recognizing that a
symplectic structure with classical dynamics is embedded in the
space of the ui representations of ccr in QFT
\cite{Vitiello:2003me} leads to show that trajectories (i.e. a
sequence of phase transitions) in such a space  may satisfy, under
convenient conditions, the criteria for chaoticity prescribed by
nonlinear classical dynamics. In a figurate way one could say that
a {\it classical blanket} covers the space of the QFT ui
representations. Moving on such a blanket describes (phase)
transitions among the representations.

The problem of the interplay between `classical and quantum' is
indeed another topic on which I will comment on in this paper and
I will show that it is intrinsic to the mathematical structure of
QFT \cite{Vitiello:2003me, Vitiello:2005aq}. The phenomenon of
decoherence in QM and the related emergence of classicality from
the quantum realm is analyzed in detail in the literature
\cite{zurek}. Similarly, although based on different formal and
conceptual frame, the emergence of macroscopic ordered patterns
and classically behaving structures out of a QFT (not QM!)
dynamics via the spontaneous breakdown of symmetry is since long
well known \cite{Umezawa,goldstone}. Examples of such classically
behaving {\it macroscopic quantum systems} are crystals,
ferromagnets, superconductors, superfluids. These are quantum
systems not in the trivial sense that they, as all other systems,
are made of quantum components, but in the sense that their
macroscopic behavior, characterized by the classical (c-number)
observable called order parameter, cannot be explained without
recourse to the underlying quantum field dynamics.

On the other hand, in recent years the problem of quantization of
a classical theory has attracted much attention in gravitation
theories and in non-hamiltonian dissipative system theories, where
a novel perspective has been proposed \cite{'tHooft:1999bx}
according to which the `emergence' of the quantum-like behavior
from a classical frame may occur. I will comment in particular on
classical deterministic systems with dissipation (information
loss) which are found to exhibit quantum behavior under convenient
conditions \cite{'tHooft:1999bx,Blasone:2000ew,Blasone:2002hq}.
The paper is organized as follows: the doubling the degrees of
freedom is discussed in Sec. 2, the two-slit experiment in Sec.
2.1, unitarily inequivalent representations in QFT in Sec. 3,
quantum dissipation in Sec. 3.1, the thermal connection and the
arrow of time in Sec. 3.2, two-mode squeezed coherent states in
Sec. 4, quantum Brownian motion in Sec. 5, the dissipative
noncommutative plane in Sec. 6. Thermal field theory in the
operatorial formalism (TFD) is presented in Sec. 7. In section 8
 the $q$--deformed Hopf algebra is shown to be a basic feature of
QFT. Entropy as a measure of entanglement  and the trajectories in
the space of the ui representations are discussed in Sec. 9 and 10
respectively. Deterministic dissipative systems are considered in
Sec. 11 with respect to the quantization problem. Section 12 is
devoted to conclusions. In this paper I have not considered the
doubling of the degrees of freedom in inflationary models and in
the problem of the quantization of the matter field in a curved
background. The interest reader is referred to the papers
\cite{Alfinito:2000bv,Martellini:1978sm,Iorio:2004bt}.


\section{Doubling the degrees of freedom}
\label{sec:2}

One of the main features underlying the QFT formalism is the
doubling of the degrees of freedom of the system under study. Such
a doubling is not simply a mathematical tool useful to describe
our system. On the contrary, it bears a physical meaning. It also
appears to be an essential feature of QM, as I will show in the
examples I am going to discuss in this paper.

The standard formalism of the density matrix \cite{schwinger,
Feynman} and of the associated  Wigner function \cite{FEStat}
suggests tha one may describe a quantum particle by splitting the
single coordinate $x(t)$ into two coordinates $x_+(t)$ (going
forward in time) and $x_-(t)$ (going backward in time). Indeed,
the standard expression for the Wigner function is \cite{FEStat},
\be \lab{W} W(p,x,t) = \frac{1}{2\pi \hbar}\int {\psi^* \left(x -
\frac{1}{2}y,t\right)\psi \left(x + \frac{1}{2}y,t\right)
e^{-i\frac{py}{\hbar}}dy} ~, \ee
where
\be \lab{1a} x_{\pm}=x\pm \frac{1}{2}y ~.\ee
By employing the Schwinger quantum operator action principle, or
recalling the mean value of a quantum operator
\be \nonumber \bar{A}(t)= (\psi (t)|A|\psi (t))= \ee
\be \nonumber  \int \! \int \psi^* (x_-,t)\, (x_-|A|x_+)\,\psi
(x_+,t)dx_+ dx_- = \ee
\be  \int \! \int (x_{+}|\rho (t)|x_{-}) (x_-|A|x_+)dx_+ dx_- .
\lab{(2)} \ee
one requires the density matrix
\be\lab{8} W(x,y,t)= (x_{+}|\rho (t)|x_{-}) = \psi^* (x_{-},t)\psi
(x_{+},t)~, \ee
to follow two copies of the Schr\"odinger equation: the forward in
time motion and the backward in time motion, respectively. These
motions  are controlled by the two Hamiltonian operators
$H_{\pm}$:
\be i\hbar {\partial \psi (x_{\pm},t) \over
\partial t}=H_{\pm}\psi (x_{\pm},t), \lab{(4a)}
\ee
%
%
%
which gives
\be i\hbar {\partial (x_+|\rho (t)|x_-) \over \partial t}= {\cal
H}\ (x_+|\rho (t)|x_-), \lab{(5a)} \ee
where
\be {\cal H}=H_+ -H_-. \lab{(5b)} \ee
Using two copies of the Hamiltonian (i.e. $H_{\pm }$) operating on
the outer product of two Hilbert spaces ${\cal F}_{+} \otimes
{\cal F}_{-}$ has been implicitly required in QM since the very
beginning of the theory. For example, from Eqs.(\ref{(5a)}),
(\ref{(5b)}) one finds immediately that the eigenvalues of ${\cal
H}$ are directly the Bohr transition frequencies $\hbar
\omega_{nm}=E_n-E_m$ which was the first clue to the explanation
of spectroscopic structure.

The notion that a quantum particle has two coordinates $x_{\pm
}(t)$ moving at the same time is therefore central
\cite{Srivastava:1995yf}.

In conclusion, the density matrix and the Wigner function  {\it
require} the introduction of a ``doubled" set of coordinates,
$(x_{\pm}, p_{\pm})$ (or $(x,p_{x})$ and $(y,p_{y})$).

Let me show how the doubling of the coordinates works in the
remarkable example of the two-slit diffraction experiment. Here I
will shortly summarize the discussion reported in
\cite{Blasone:1998xt}.

\subsection{The two-slit experiment}
\label{sec:2a1}

In order to derive the diffraction pattern it is required to know
the wave function $\psi_0(x)$ of the particle when it ``passes
through the slits'' at time zero. In other words, one searches for
the density matrix
\be (x_+|\rho_0|x_-)=\psi^*_0 (x_-)\psi_0 (x_+). \lab{(11)} \ee
The probability density for the electron to be found at position
$x$ at the detector screen at a later time $t$ is written as
\be P(x,t)=(x|\rho (t)|x)=\psi^* (x,t)\psi (x,t)  \lab{(12)} \ee
in terms of the  solution $\psi (x,t)$ to the free particle
Schr\"odinger equation
\be \psi (x,t)=\Big({M\over 2\pi\hbar it} \Big)^{1/2}
\int_{-\infty}^{\infty} e^{[\frac{i}{\hbar } A(x-x^\prime,t)
]}\psi_0 (x^\prime ) dx^\prime ,\lab{(13a)} \ee
where
\be A(x-x^\prime,t)={M(x-x^\prime )^2\over 2t} \lab{(13b)} \ee
is the Hamilton-Jacobi action for a classical free particle to
move from $x^\prime $ to $x$ in a time $t$.  Eqs.
(\ref{(11)})-(\ref{(13b)}) then imply that
\be \lab{(14)}
 P(x,t)=  {M\over 2\pi\hbar t}
\int_{-\infty}^{\infty}\int_{-\infty}^{\infty}
e^{\le[iM{(x-x_+)^2-(x-x_-)^2 \over 2\hbar t} \right]}
(x_+|\rho_0|x_-) dx_+dx_-. \ee
Eq.(\ref{(14)}) shows that  $P(x,t)$ would not oscillate in $x$,
i.e. there would not be the usual quantum diffraction, if $x_+ =
x_-$. In Eq.(\ref{(14)}), in order to have quantum interference
the forward in time action $A(x-x_+,t)$ must be different from the
backward in time action $A(x-x_-,t)$: the non-trivial dependence
of the density matrix $(x_+|\rho_0|x_-)$ when the electron
``passes through the slits'' on the difference $(x_+-x_-)$
crucially determines the quantum nature of the phenomenon.

\begin{figure}[htbp]
\begin{center}
\mbox{\epsfig{file=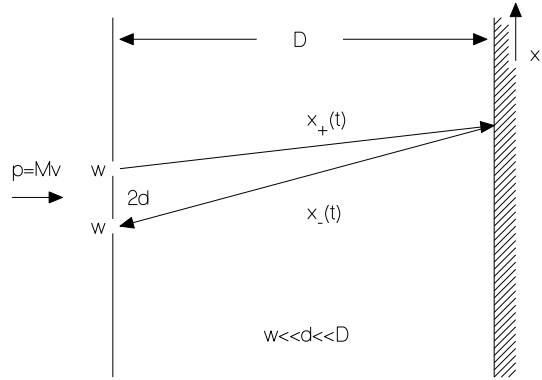,width=7.0cm}} \caption{Two slit
experiment.} \label{fig:1}
\end{center}
\end{figure}

%
%
%
%





In the quantum diffraction experiment the experimental apparatus
is prepared so that $w \ll d \ll D$, with $w$ the opening of the
slits which are separated by a distance $2d$. $D$ is the distance
between the slits and the screen (Fig.1). The diffraction pattern
is described by $|x|\gg|x_\pm|$. By defining $K={Mvd \over \hbar
D},\ \ \beta={w\over d}$, with $v=D/t$ the velocity of the
incident electron,  Eq.(\ref{(14)}) leads \cite{Blasone:1998xt} to
the usual result
\be P(x,D)\approx {4\over \pi \beta K x^2} \; \cos^2(Kx)
\sin^2(\beta Kx), \lab{(21)} \ee
where the initial wave function
\be \psi_0(x)={1\over \sqrt{2}}\Big[\phi (x-d)+\phi (x+d)\Big],
\lab{(17)} \ee
with $\phi(x)={1\over \sqrt{w}}$  if  $|x|\leq \frac{w}{2}$ and
zero otherwise, has been used. From Eqs.(\ref{(11)}) and
(\ref{(17)}) we have
\bea \nonumber (x_+|\rho_0|x_-)= && {1\over 2}\Big\{\phi
(x_+-d)\phi (x_--d)+ \phi (x_++d) \phi (x_-+d) \\ \lab{(19)} &&
+\phi (x_+-d)\phi (x_-+d) +\phi (x_++d)\phi (x_--d)\Big\}.
  \eea
In the rhs of Eq.(\ref{(19)}) the first and the second terms
describe the classical processes of the particle going forward and
backward in time through slit 1 and going forward and backward in
time through slit 2, respectively. In these processes it is
$x_+(t)=x_-(t)$ and in such cases no diffraction is observed on
the screen. The third term and the fourth term describe the
particle going forward in time through slit 1 and backward in time
through slit 2, or forward in time through slit 2 and backward in
time through slit 1, respectively. These are the terms generating
quantum interference since $|x_+(t)-x_-(t)|>0$.

In conclusion, the doubling the system coordinates, $x(t)
\rightarrow (x_+(t),x_-(t))$ plays a crucial r\^ole in the
description of the quantum system. If $x(t)\equiv x_{+}(t)\equiv
x_{-}(t)$, then the system behavior appears to be a classical one.
When forward in time and backward in time motions are (at the same
time) unequal $x_+(t)\ne x_-(t)$, then the system is behaving in a
quantum mechanical fashion and interference patterns appear in
measured position probability densities.

I will not comment further on  the two-slit experiment. In the
following Section I go back to the general discussion of the
doubling of the degrees of freedom and of its meaning in QFT.

\section{Unitarily inequivalent representations in QFT}
\label{sec:3}

The mathematical r\^ole and the physical meaning of the doubling
of the degrees of freedom fully appears in dealing with phase
transitions, with equilibrium and non-equilibrium thermal field
theories and with dissipative, open systems. In these cases the
doubling of the degrees of freedom appears to be a structural
feature of QFT since it strictly relates with the existence of the
unitarily inequivalent representations of the ccr in QFT.

Let me consider the case of dissipation \cite{Celeghini:1992yv,
FE,Blasone:1996yh}. I will discuss the canonical quantization of
the damped (simple) harmonic oscillator (dho), which is a simple
prototype of dissipative system.

\subsection{Quantum dissipation}
\label{sec:3a}

Dissipation enters into our considerations if there is a coupling
to a thermal reservoir yielding a mechanical resistance
\begin{math} R \end{math}. According to the discussion in Section 2,
the equation of motion for the density matrix is given by Eq.
(\ref{(5a)}), where now the Hamiltonian \begin{math} {\cal H}
\end{math} for motion in the
\begin{math} (x_+,x_-) \end{math} plane is
\cite{Srivastava:1995yf,Blasone:1998xt}
\be \nonumber {\cal H} = \frac{1}{2M} (p_+ - \frac{R}{2}\, x_-
)^2-\frac{1}{2M} (p_- + \frac{R}{2}\, x_+ )^2
 +U(x_+)-U(x_-) ~, \lab{(7ba)}
\ee
where $p_\pm =-i\hbar {\partial \over \partial x_\pm}$. In order
to simplify the discussion, it is convenient, without loss of
generality, to make an explicit (simple) choice for $U(x_{\pm})$,
say $U(x_{\pm}) = \frac{1}{2}\kappa x^{2}_{\pm}$. By choosing as
doubled coordinates the pair $(x,y)$ with
\be \lab{1ay} y = x_{+}- x_{-} ~,\ee
the Hamiltonian (\ref{(7ba)}) can be derived from the Lagrangian
(see \cite{Celeghini:1992yv} - \cite{bateman})
\be\lab{(2.2)} L = M \dot x \dot y + \half R ( x \dot y - \dot x y
) - \kappa x y~. \quad  \ee
The  system described by  (\ref{(2.2)}) is sometimes called
Bateman's dual system \cite{bateman}. I observe that the doubling
imposed by the density matrix and the Wigner function formalism,
as seen in Section 2, here finds its physical justification in the
fact that the canonical quantization scheme can only deal with an
isolated system. In the present case our system has been assumed
to be coupled with a thermal reservoir and it is then necessary to
{\it close} the system by including the reservoir. This is
achieved by doubling the phase-space dimensions
\cite{Celeghini:1992yv, FE}. Eq. (\ref{(2.2)}) is indeed the
closed system Lagrangian.

By varying Eq. (\ref{(2.2)}) with respect to $y$ gives
\be\lab{(2.1)} M \ddot x + R \dot x + \kappa x = 0 \quad , \ee
whereas variation with respect to $x$ gives
\be\lab{(2.3)} M \ddot y - R \dot y + \kappa y = 0 \quad , \ee
which is the {\it time reversed} ($R \rightarrow - R$) of Eq.
(\ref{(2.1)}). The physical meaning of the doubled degree of
freedom $y$  is now manifest: $y$ may be thought of as describing
an effective degree of freedom for the reservoir to which the
system (\ref{(2.1)}) is coupled. The canonical momenta are  given
by ${\, p_{x} \equiv {{\partial L}\over{\partial \dot x}} = M \dot
y - \half R y}$ ; ${p_{y} \equiv {{\partial L}\over{\partial \dot
y}} = M \dot x + \half R x}$.  For a discussion of Hamiltonian
systems of this kind see also \cite{Tsue:1994nz,Banerjee:2001yc}.
Canonical quantization is performed by introducing the commutators
\be\lab{(2.3a)} [ x , p_{x} ]= i\, \hbar = [ y , p_{y} ] , ~~~ [ x
, y ] = 0 = [ p_{x} , p_{y} ] ~, \ee
and the corresponding sets of annihilation and creation operators
\be\lab{(2.5)} \alpha  \equiv \left ({1\over{2 \hbar \Omega}}
\right )^{1\over{2}} \left ( {{p_{x}}\over{\sqrt{M}}} - i \sqrt{M}
\Omega x \right ) , \quad \alpha^{\dagger} \equiv \left ({1\over{2
\hbar \Omega}} \right )^{1\over{2}} \left (
{{p_{x}}\over{\sqrt{M}}} + i \sqrt{M} \Omega x \right ) , \ee

\be\lab{(2.5a)} \beta \equiv \left ({1\over{2 \hbar \Omega}}
\right )^{1\over{2}} \left ( {{p_{y}}\over{\sqrt{M}}} - i \sqrt{M}
\Omega y \right ) , \quad \beta^{\dagger} \equiv \left ({1\over{2
\hbar \Omega}} \right )^{1\over{2}} \left (
{{p_{y}}\over{\sqrt{M}}} + i \sqrt{M} \Omega y \right ) , \ee

\be\lab{(2.5b)} [\, \alpha , \alpha^{\dagger} \,] = 1 = [\, \beta
, \beta^{\dagger} \, ] \quad , \quad  [\, \alpha , \beta \,] = 0 =
[\, \alpha , \beta^{\dagger} \, ] \quad . \ee \noindent I have
introduced ${\Omega \equiv \left [ {1\over{M}} \left ( \kappa -
{{R^{2}}\over{4 M}} \right ) \right ]^{1\over{2}}}$, the common
frequency of the two oscillators Eq. (\ref{(2.1)}) and Eq.
(\ref{(2.3)}), assuming $\Omega$ real, hence ${ \kappa >
{{R^{2}}\over{4 M}}}$ (case of no overdamping).

In Section 5 I show that, at quantum level, the $\beta$ modes
allow quantum noise effects arising from the imaginary part of the
action \cite{Srivastava:1995yf}. Moreover, in Section 8 the modes
$\alpha$ and $\beta$ will be shown to be the modes involved in the
coproduct operator of the underlying $q$--deformed Hopf algebra
structure. The $q$-deformation parameter turns out to be a
function of $R$, $M$ and $t$.

By using the canonical linear transformations $ {A \equiv
{1\over{\sqrt 2}} ( \alpha + \beta )} ,$~ $ {B \equiv
{1\over{\sqrt 2}} ( \alpha - \beta )},~ $ the quantum Hamiltonian
 $H$ is then obtained \cite{Celeghini:1992yv,FE} as
\be\lab{(2.71)}
 H =  H_{0} +  H_{I} \quad ,
\ee
\be\lab{(2.7)}
 H_{0} = \hbar \Omega ( A^{\dagger} A - B^{\dagger} B ) \quad , \quad
H_{I} = i \hbar \Gamma ( A^{\dagger} B^{\dagger} - A B ) \quad ,
\ee
where the decay constant for the classical variable $x(t)$ is
denoted by $\Gamma \equiv {{R}\over{2 M}}$.

In conclusion, {\it the states generated by $B^{\dagger}$
represent the sink where the energy dissipated by the quantum
damped oscillator flows: the $B$-oscillator represents the
reservoir or heat bath coupled to the $A$-oscillator}.

The dynamical group structure associated with the system of
coupled quantum oscillators is that of $SU(1,1)$. The two mode
realization of the algebra $su(1,1)$ is indeed generated by $
J_{+} = A^{\dagger} B^{\dagger} , \quad J_{-} = J_{+}^{\dagger} =
A B , \quad J_{3} = {1\over{2}} (A^{\dagger} A + B^{\dagger} B +
1) ,~ $ $ [\, J_{+} , J_{-}\, ] = - 2 J_{3} , \quad [\, J_{3}  ,
J_{\pm}\, ] = \pm J_{\pm} .$ The Casimir operator ${\cal C}$ is
${{\cal C}^{2} \equiv {1\over {4}} + J_{3}^{2} - {1\over{2}} (
J_{+} J{-} + J_{-} J_{+} )}$ ${= {1\over{4}} ( A^{\dagger} A -
B^{\dagger} B)^{2}} $.

I also observe that $ [\, H_{0} , H_{I}\, ] = 0 $. The time
evolution of the vacuum $ |0> \equiv | n_{A} = 0 , n_{B} = 0 >  =
|0> \otimes ~|0> ~,~ (A \otimes ~1) |0> \otimes ~|0> \equiv A|0> =
0; ~ (1 \otimes ~B) |0> \otimes ~|0> \equiv B|0> = 0 $, is
controlled by $H_{I}$
$$
|0(t)> = \exp{ \left ( - i t {H \over{\hbar}} \right )} |0> =
\exp{ \left ( - i t {H_{I} \over{\hbar}} \right )} |0>
$$
\be\lab{(2.8a)} = {1\over{\cosh{(\Gamma t)}}} \exp{ \left ( \tanh
{(\Gamma t)} A^{\dagger} B^{\dagger} \right )}|0> \quad , \ee

\be\lab{(2.8b)} <0(t) | 0(t)> = 1~ \quad \forall t~, \ee
\be\lab{(2.9)} \lim_{t\to \infty} <0(t) | 0> \, \propto \lim_{t\to
\infty} \exp{( - t  \Gamma )} = 0 \quad . \ee

Notice that once one sets the initial condition of positiveness
for the eigenvalues of $H_{0}$, such a condition is preserved by
the time evolution since $H_{0}$ is the Casimir operator (it
commutes with $H_{I}$). In other words, there is no danger of
dealing with energy spectrum unbounded from below. Time evolution
for creation and annihilation operators is given by
\be\lab{(2.10a)} A \mapsto A(t) = {\rm e}^{- i {t\over{\hbar}}
H_{I}} A ~{\rm e}^{i {t\over{\hbar}} H_{I}} = A \cosh{(\Gamma t)}
- B^{\dagger} \sinh{(\Gamma t)} ~, \ee \be \lab{(2.10b)} B \mapsto
B(t) = {\rm e}^{- i {t\over{\hbar}} H_{I}} B ~{\rm e}^{i
{t\over{\hbar}} H_{I}} = B \cosh{(\Gamma t)} - A^{\dagger}
\sinh{(\Gamma t)} \quad \ee
and h.c.. I note that Eqs. (\ref{(2.10a)}) and (\ref{(2.10b)}) are
Bogolubov transformations: they are canonical transformations
preserving the ccr. Eq. (\ref{(2.9)}) expresses the instability
(decay) of the vacuum under the evolution operator $\exp{ \left (
- i t {H_{I} \over{\hbar}} \right )}$. In other words, time
evolution leads out of the Hilbert space of the states. {\it This
means that the QM framework is not suitable for the canonical
quantization of the damped harmonic oscillator}.  A way out from
such a difficulty is provided by QFT \cite{Celeghini:1992yv}:  the
proper way to perform the canonical quantization of the dho turns
out to be working in the framework of QFT. In fact, for many
degrees of freedom the time evolution operator ${\cal U}(t)$ and
the vacuum are formally (at finite volume) given by
\be \lab{(2.11)} {\cal U}(t) =\prod_{\kappa}{\exp{\Bigl (
\Gamma_{\kappa} t \bigl ( A_{\kappa}^{\dagger}
B_{\kappa}^{\dagger} - A_{\kappa} B_{\kappa} \bigr ) \Bigr )}},
\ee
\be\lab{(2.12)} |0(t)> = \prod_{\kappa}
{1\over{\cosh{(\Gamma_{\kappa} t)}}} \exp{ \left ( \tanh
{(\Gamma_{\kappa} t)} A_{\kappa}^{\dagger} B_{\kappa}^{\dagger}
\right )} |0> \quad , \ee
with $<0(t) | 0(t)> = 1, ~ \forall t $~. Using the continuous
limit relation $ \sum_{\kappa} \mapsto {V\over{(2 \pi)^{3}}} \int
\! d^{3}{\kappa}$, in the infinite-volume limit we have (for $\int
\! d^{3} \kappa ~ \Gamma_{\kappa}$ finite and positive)
\be\lab{(2.13)} {<0(t) | 0> \rightarrow 0~~ {\rm as}~~
V\rightarrow \infty } ~~~\forall~  t~  , \ee
and in general, $ {<0(t) | 0(t') > \rightarrow 0~ {\rm as}~
V\rightarrow \infty} ~~\forall~t$~ and~ $t' ,~ t' \neq t. $ At
each time {\it t} a representation $\{ |0(t)> \}$ of the ccr is
defined and turns out to be ui to any other representation $\{
|0(t')>,~\forall t'\neq t \}$ in the infinite volume limit. In
such a way the quantum dho evolves in time through ui
representations of ccr ({\it tunneling}). I remark that  $| 0(t)>$
is a two-mode time dependent generalized coherent state
\cite{22.,perelomov}. Also note that
\be \lab{(2.13a)} {\cal N}_{A_{\kappa}}(t) = <0(t)
|A_{\kappa}^{\dagger} A_{\kappa}| 0(t)> =
 \sinh^{2} \Gamma t ~ , \ee

The Bogolubov transformations, Eqs. (\ref{(2.10a)}) and
(\ref{(2.10b)}) can be implemented for every $\kappa$ as inner
automorphism for the algebra ${su(1,1)}_{\kappa}$. At each time
{\it t} one has a copy $\{ A_{\kappa}(t) , A_{\kappa}^{\dagger}(t)
, B_{\kappa}(t) , B_{\kappa}^{\dagger}(t) \, ; \, | 0(t) >\, |\,
\forall {\kappa} \}$ of the original algebra induced by the time
evolution operator which can thus be thought of as a generator of
the group of automorphisms of ${\bigoplus_{\kappa}
su(1,1)_{\kappa}}$ parameterized by time {\it t} (we have a
realization of the operator algebra at each time t, which can be
implemented by Gel'fand-Naimark-Segal construction in the
C*-algebra formalism \cite{Bratteli,ojima}). Notice that the
various copies become unitarily inequivalent in the
infinite-volume limit, as shown by Eqs. (\ref{(2.13)}): the space
of the states splits into ui representations of the ccr each one
labeled by time parameter {\it t}. As usual, one works at finite
volume and only at the end of the computations the limit $V \to
\infty$ is performed.

Finally, I note that the ``negative" kinematic term in the
Hamiltonian (\ref{(2.7)}) (or (\ref{(7ba)})) also appears in
two-dimensional gravity models where, in general, two different
strategies are adopted in the quantization procedure \cite{JA}:
the Schr\"odinger representation approach, where no negative norm
appears, and the string/conformal field theory approach where
negative norm states arise as in Gupta-Bleurer electrodynamics.

\subsection{The thermal connection and the arrow of time}
\label{sec:3b}

It is useful \cite{Celeghini:1992yv} to introduce the  functional
${\cal F}_{A}$ for the $A$-modes
\be\lab{(2.14)} {\cal F}_{A} \equiv <0(t)| \Bigl ( H_{A} -
{1\over{\beta}} S_{A} \Bigr ) |0(t)> \quad , \ee
where $\beta$ is a non-zero c-number, $H_{A}$ is the part of
$H_{0}$ relative to $A$- modes only, namely $H_{A} = \sum_{\kappa}
\hbar \Omega_{\kappa} A_{\kappa}^{\dagger} A_{\kappa}$, and the
$S_{A}$ is given by
\be\lab{(2.15)}
 S_{A} \equiv - \sum_{\kappa} \Bigl \{ A_{\kappa}^{\dagger} A_{\kappa}
\ln \sinh^{2} \bigl ( \Gamma_{\kappa} t \bigr ) - A_{\kappa}
A_{\kappa}^{\dagger} \ln \cosh^{2} \bigl ( \Gamma_{\kappa} t \bigr
) \Bigr \} \quad . \ee
One then considers the extremal condition ${{\partial {\cal
F}_{A}}\over{\partial \vartheta_{\kappa}}} = 0 \quad
 \forall \kappa ~,~\vartheta_{\kappa} \equiv \Gamma_{\kappa} t$~
to be satisfied in each representation, and using the definition
$E_{\kappa} \equiv \hbar \Omega_{\kappa}$, one finds
\be
{\cal N}_{A_{\kappa}}(t) = \sinh^{2} \bigl ( \Gamma_{\kappa} t
\bigr ) = {1\over{{\rm e}^{\beta (t) E_{\kappa}} - 1}} \quad ,
\lab{(2.16)} \ee
which is the Bose distribution for $A_{\kappa}$ at time {\it t},
{\it provided} $\beta (t)$ is the (time-dependent) inverse
temperature. Inspection of Eqs.  (\ref{(2.14)}) and (\ref{(2.15)})
then suggests that ${\cal F}_{A}$ and $S_{A}$  can be interpreted
as the {\it free energy} and the {\it entropy}, respectively. I
will comment more about this in Section 7 and 9.

$\{ |0(t)> \}$ is  thus recognized to be a representation of the
ccr at finite temperature (it turns out to be equivalent to the
thermo field dynamics (TFD) representation $\{ |0(\beta)> \}$~
\cite{Umezawa,TFD}, see Section 7). Use of Eq. (\ref{(2.15)})
shows that
\be\lab{()} {{\partial}\over{\partial t}} |0(t)> =  - \left (
{1\over{2}} {{\partial {\cal S}}\over{\partial t}} \right ) |0(t)>
\quad . \lab{(2.17)} \ee
One thus see that $i \left ( {1\over{2}} \hbar {{\partial {\cal
S}}\over{\partial t}} \right )$ is the generator of time
translations, namely time evolution is controlled by the entropy
variations \cite{DeFilippo:1977bk}. It is remarkable that the same
dynamical variable ${\cal S}$ whose expectation value is formally
the entropy also controls time evolution: damping (or, more
generally, dissipation) implies indeed the choice of a privileged
direction in time evolution ({\it arrow of time}) with a
consequent breaking of time-reversal invariance.

One may also show that $ d {\cal F}_{A} = d E_{A} -
{1\over{\beta}} d {\cal S}_{A}=0~, $ which expresses the first
principle of thermodynamics for a system coupled with environment
at constant temperature and in absence of mechanical work. As
usual, one may define  heat as ${dQ={1\over{\beta}} dS}$ and see
that the change in time $d {\cal N}_{A}$ of particles condensed in
the vacuum turns out into heat dissipation $dQ$:
\be \lab{dq} d E_{A} = \sum_{\kappa} \hbar \Omega_{\kappa}
\dot{{\cal N}}_{A_{\kappa}}(t)d t = {1\over{\beta}} d S = d Q ~.
\ee
Here  ${\dot{\cal N}}_{A_{\kappa}}$ denotes the time derivative of
${\cal N}_{A_{\kappa}}$.

It is interesting to observe that the thermodynamic arrow of time,
whose direction is defined by the increasing entropy direction,
points in the same direction of the cosmological arrow of time,
namely the inflating time direction for the expanding Universe.
This can be shown by considering indeed the quantization of
inflationary models \cite{Alfinito:2000bv} (see also
\cite{Martellini:1978sm}). The concordance between the two arrows
of time (and also with the psychological arrow of time, see refs.
\cite{Vitiello:1995wv}) is not at all granted and is a subject of
an ongoing debate (see, e.g., \cite{Hawking:1996ny}).

In Section 6 I will show that quantum dissipation induces a
dissipative phase interference \cite{Blasone:1998xt}, analogous to
the Aharonov-Bohm phase \cite{Berry}, and a noncommutative
geometry in the plane $(x_{+}, x_{-})$ \cite{noncomm}.

The quantum dissipation Lagrangian model discussed above is
strictly related with the squeezed coherent states in quantum
optics and with the quantum Brownian motion. I will briefly
discuss these two topics in follolwing Sections.

\section{Two-mode squeezed coherent states}
\label{sec:4a}

Here I will only mention that  in the quantum damped oscillator
treatment presented above the time evolution operator ${\cal
U}(t)$ written in terms of the $\alpha$ and $\beta$ modes (Eqs.
(\ref{(2.5)}) and (\ref{(2.5a)})) is given by
\bea  \nonumber {\cal U}(t) \equiv \exp{\left ( -i t {{H_{I}}
\over {\hbar}}\right )} &=& \prod_{\kappa}{\exp\Bigl(-{
\theta_{\kappa}\over{ 2}}\bigl( {\alpha}_{\kappa}^2 -
{\alpha}_{\kappa}^{\dagger 2}\bigr) \Bigr)
\exp\Bigl({\theta_{\kappa} \over{ 2}}\bigl( {\beta}_{\kappa}^2 -
{\beta}_{\kappa}^{\dagger 2}\bigr) \Bigr)} \\
\lab{(25)} &\equiv& {{\prod}_{\kappa}} {\hat S}_{\al}(
\theta_{\kappa} ){\hat S}_{\beta} (-\theta_{\kappa}  ) ~ , \eea
with $\hat {S_{\al}}( \theta_{\kappa}  ) {\equiv} \exp\bigl(-{{
{\theta}_{\kappa}  }\over{2}}\bigl({\al_{\kappa}}^{2}
-{\al_{\kappa}}^{ {\dagger} 2}\bigr)\bigr)$ and similar expression
for ${\hat {S_{\beta}}}( -{\theta}_{\kappa})$ with ${\beta}$ and
${\beta}^{\dagger}$ replacing $\al$ and $\al^{\dagger}$,
respectively. The operators ${\hat S}_{\al}({ \theta}_{\kappa}  )$
and ${\hat S}_{\beta}( -{\theta}_{\kappa})$ are the squeezing
operators for the $\al_{\kappa}$ and the ${\beta}_{\kappa}$ modes,
respectively, as well known in quantum optics \cite{[32]}. The set
$\theta \equiv
 \{ \theta_{\kappa} \equiv  \Gamma_{\kappa} t \}$ as well as each
 $ \theta_{\kappa} $ for all $\kappa$ is called the squeezing parameter.
The state $|0(t)>$  is thus a squeezed coherent states at each
time $t$.

To illustrate the effect of the squeezing, let me focus the
attention only on the $\alpha_{\kappa}$ modes for sake of
definiteness. For the $\beta$ modes one can proceed in a similar
way. As usual, for given $\kappa$ I express the $\al$ mode in
terms of conjugate variables of the corresponding oscillator. By
using dimensionless quantities I thus write $\al = X + iY$, with
$[X,Y] = {i\over{2}}$. The uncertainty relation is ${\Delta }X
{\Delta}Y = {1\over{4}}$ , with ${{\Delta }X}^{2} =
{{\Delta}Y}^{2} = {1\over{4}}$  for (minimum uncertainty) coherent
states. The squeezing occurs when ${{\Delta }X}^{2} < {1\over{4}}$
{\it and} ${{\Delta}Y}^{2} > {1\over{4}}$ (or ${{\Delta }X}^{2}
> {1\over{4}}$ {\it and} ${{\Delta}Y}^{2} < {1\over{4}}$) in such a way
that the uncertainty relation remains unchanged. Under the action
of ${\cal U}(t)$ the variances ${\Delta }X$ and ${\Delta}Y$ are
indeed squeezed as
\be \lab{(26a)}{{\Delta }{X}}^{2}(\theta )
  = {{\Delta }
 {X}}^{2}\exp(2{\theta})~~,~~~~
  {{\Delta }{Y}}^{2}(\theta )
   =   {{\Delta }
  {Y}}^{2}\exp(-2{\theta} )~~.
\ee
For the tilde-mode similar relations are obtained for the
corresponding variances, say $\tilde X$ and $\tilde Y$:
\be \lab{(26b)}
{{\Delta }{\tilde X}}^{2}(\theta )
  = {{\Delta }
 {\tilde X}}^{2}\exp(-2{\theta})~~,~~~~
  {{\Delta }{\tilde Y}}^{2}(\theta )
   =   {{\Delta }
  {\tilde Y}}^{2}\exp(2{\theta} )~~.
\ee
For positive $\theta$, squeezing then reduces the variances of the
$Y$ and $\tilde X$ variables, while the variances of the $X$ and
$\tilde Y$ variables grow by the same amount so to keep the
uncertainty relations unchanged. This reflects, in terms of the
$A$ and $B$ modes, the constancy of the difference ${\cal
N}_{A_{\kappa}} - {\cal N}_ {B_{\kappa}}$ against separate, but
equal, changes of ${\cal N}_{A_{\kappa}}$ and ${\cal
N}_{B_{\kappa}}$ (degeneracy of the states $|0(t)>$ {\it labelled}
by different ${\cal N}_{A_{\kappa}}$, or different ${\cal
N}_{B_{\kappa}}$, cf. Eq. (\ref{(2.13a)}).

In conclusion, the  $\theta$-set $\{ \theta_{\kappa}({\cal
N}_{\kappa}) \}$, is nothing but the squeezing parameter
classifying the squeezed coherent states in the hyperplane
$(X,{\tilde X} ; Y, {\tilde Y})$. Note that to different squeezed
states (different  $\theta$-sets) are associated unitarily
inequivalent representations of the ccr's in the infinite volume
limit. Also note that in the limit $t \rightarrow  \infty$ the
variances of the variables $Y$ and ${\tilde X}$ become infinity
making them completely spread out.

Further details on the squeezing states and their relation with
deformed algebraic structures in QFT can be found in refs.
\cite{Celeghini:1989qc,Iorio:1994jk,Celeghini:1991jw}.

\section{Quantum Brownian motion}
\label{sec:4}

By following Schwinger \cite{schwinger}, the description of a
Brownian particle of mass $M$ moving in a potential U(x) with a
damping resistance R, interacting with a thermal bath at
temperature $T$ is provided by \cite{Srivastava:1995yf,
Blasone:1998xt}
\be \lab{(7a)} {\cal H}_{Brownian}= {\cal H} - {ik_BTR \over
\hbar}(x_+-x_-)^2~. \ee
Here ${\cal H}$ is given by Eq. (\ref{(7ba)}) and the evolution
equation for the density matrix is
\be i\hbar {\partial (x_+|\rho (t)|x_-) \over
\partial t} =
{\cal H}( x_+ \left| \rho (t) \right| x_- ) - ( x_+ \left|N[\rho ]
\right| x_- )~, \lab{(7c)} \ee
where  $N[\rho ]\approx (ik_BTR/\hbar )[x,[x,\rho ]]$ describes
the effects of the reservoir random thermal noise
\cite{Srivastava:1995yf,Blasone:1998xt}.

In general the density operator in the above expression describes
a mixed statistical state. The thermal bath contribution to the
right hand side of Eq.(\ref{(7a)}), proportional to fluid
temperature T, can be shown \cite{Blasone:1998xt} to be equivalent
to a white noise fluctuation source coupling the forward and
backward motions according to
\be <y(t)y(t^\prime )>_{noise}={\hbar^2 \over 2Rk_BT}\delta
(t-t^\prime ), \lab{(8)} \ee
so that thermal fluctuations are always occurring in the
difference $y = x_+ -x_- $ between forward in time and backward in
time coordinates.

The correlation function for the random force $f$  on the particle
due to the bath is given by $ G(t-s)=(i/\hbar )<f(t)f(s)>. $ The
retarded and advanced Greens functions are studied in ref.
\cite{Srivastava:1995yf} and for brevity I omit here their
discussion. The mechanical resistance is defined by $R=lim_{\omega
\rightarrow 0}{\cal R}e Z(\omega +i0^+), $ ~with the mechanical
impedance $Z(\zeta )$ (analytic in the upper half complex
frequency plane ${\cal I}m ~\zeta >0$) determined by the retarded
Greens function $ -i\zeta Z(\zeta )=\int_0^\infty dt
G_{ret}(t)e^{i\zeta t}. $ The time domain quantum noise in the
fluctuating random force is $ N(t-s)=(1/2)<f(t)f(s)+f(s)f(t)>. $

The interaction between the bath and the particle is evaluated by
following Feynman and Vernon and one finds
\cite{Srivastava:1995yf} for the real and the imaginary part of
the action
\be\lab{(3.10a)} {\cal R}e{\cal A}[x,y]=\int_{t_i}^{t_f}dt L, \ee
\be\lab{(3.10c)} {\cal I}m{\cal A}[x,y]= (1/2\hbar
)\int_{t_i}^{t_f}\int_{t_i}^{t_f}dtdsN(t-s)y(t)y(s), \ee
respectively, where $L$ is defined in Eq. (\ref{(2.2)}) for the
given choice of $U(x_{\pm})$ there adopted (without loss of
generality).

I observe that at the classical level the ``extra'' coordinate
$y$, is usually constrained to vanish. Note that $y(t)=0$ is a
true solution to Eqs. (\ref{(2.3)}) so that the constraint is {\it
not} in violation of the equations of motion. From Eqs.
(\ref{(3.10a)}) and (\ref{(3.10c)}) one sees that {\it at quantum
level nonzero $y$ allows quantum noise effects arising from the
imaginary part of the action}. On the contrary, in the classical
``$\hbar \rightarrow 0$'' limit nonzero $y$ yields an ``unlikely
process'' in view of the large imaginary part of the action
implicit in Eq. (\ref{(3.10c)}). Thus, the meaning of the
constraint $y=0$ at the classical level is the one of avoiding
such ``unlikely process''.

The r\^ole of the doubled $y$ coordinate (the quantum $\beta$, or
$B$ mode in the discussion of the previous Section) is thus shown
again to be absolutely crucial in the quantum regime. There it
accounts for the quantum noise in the fluctuating random force in
the system-environment coupling \cite{Srivastava:1995yf}: in the
limit of $y \rightarrow 0$ (i.e. for $x_{+} = x_{-}$) quantum
effects are lost and the classical limit is obtained.

It is interesting to remark that the forward and backward in time
velocity  components \begin{math} v_\pm =\dot{x}_\pm
\end{math} in the \begin{math} (x_+,x_-) \end{math} plane
\be v_{\pm }={\partial {\cal H} \over
\partial p_{\pm }} =\pm \, \frac{1}{M}( p_\pm \mp \frac{R}{2}\,
x_\mp) \lab{(9)} \ee
do not commute
\be [v_+,v_-]=i\hbar \,{R\over M^2},  \lab{(10)} \ee
and it is thus impossible to fix these velocities  $v_+$ and $v_-$
as being identical. Eq.(\ref{(10)}) is similar to the usual
commutation relations for the quantum velocities ${\bf v}=({\bf
p}-(e{\bf A}/c))/M$ of a charged particle moving in a magnetic
field ${\bf B}$; i.e. $[v_1,v_2]=(i\hbar eB_3/M^2c)$. Just as the
magnetic field ${\bf B}$ induces an Aharonov-Bohm phase
interference for the charged particle, the Brownian motion
friction coefficient $R$ induces an analogous phase interference
between forward and backward motion which expresses itself as
mechanical damping. Eq. (\ref{(10)}) will be also discussed in
connection with noncommutative geometry induced by quantum
dissipation \cite{noncomm}. I will comment more on this in the
next Section.

In the discussion above I have considered the low temperature
limit: $T \ll T_\gamma $ where $k_BT_\gamma=\hbar \gamma={\hbar
R\over 2M}$. At high temperature,  $T \gg T_\gamma $, the thermal
bath motion suppresses the probability for $x_+\ne x_-$ due to the
thermal term $(k_BTR /\hbar)(x_+-x_-)^2$ in Eq.(\ref{(7a)}) (cf.
also Eq. (\ref{(8)})). By writing the diffusion coefficient
$D={k_BT\over R}$ as
\be D={T\over T_\gamma }\Big({\hbar \over 2M}\Big), \lab{(54)} \ee
the condition for classical Brownian motion for high mass
particles is that $D \gg (\hbar /2M)$, and the condition for
quantum interference with low mass particles is that $D \ll (\hbar
/2M)$.  In colloidal systems, for example, classical Brownian
motion for large particles would appear to dominate the motion. In
a fluid at room temperature it is typically $D\sim (\hbar /2M)$
for a single atom, or, equivalently, $T\sim T_\gamma $, so that
the r\^ole played by quantum mechanics, although perhaps not
dominant, may be an important one in the Brownian motion.

\section{Dissipative noncommutative plane}
\label{sec:4}

The harmonic oscillator on the noncommutative plane, the motion of
a particle in an external magnetic field and the Landau problem on
the noncommutative sphere are only few examples of systems whose
noncommutative geometry has been studied in detail. Noncommutative
geometries are also of interest in Cern-Simons gauge theories, the
usual gauge theories and string theories, in gravity theory
\cite{Dunne:1989hv,Aschieri:2005yw}. Here I show that quantum
dissipation induces noncommutative geometry in the
\begin{math} (x_+,x_-) \end{math} plane \cite{noncomm}.

The velocity components \begin{math} v_\pm =\dot{x}_\pm \end{math}
in the \begin{math} (x_+,x_-) \end{math} plane are given Eq.
(\ref{(9)}). Similarly,
\begin{equation}
\dot{p}_\pm =- \frac{\partial {\cal H}}{\partial x_\pm } =\mp
U^\prime (x_\pm )\mp \frac{Rv_\mp }{2}\ . \label{QD11}
\end{equation}
From Eqs.(\ref{(9)}) and (\ref{QD11}) it follows that
\begin{equation}
M\dot{v}_\pm +Rv_{\mp}+U^\prime (x_\pm )=0 ~.\label{QD12}
\end{equation}
When the choice $U(x_{\pm}) = \frac{1}{2}\kappa x^{2}_{\pm}$ is
made, these are equivalent to the equations Eq.(\ref{(2.1)}) and
(\ref{(2.3)}).  The classical equation of motion including
dissipation thereby holds true if
\begin{math} x_+(t)\approx x_-(t)\approx x(t)\end{math}:
\begin{equation}
M\dot{v} + Rv + U^\prime (x)=0 ~.\label{QD13}
\end{equation}

If one defines
\begin{equation}
Mv_\pm =\hbar K_\pm , \label{momentum}
\end{equation}
then  Eq.(\ref{(10)}) gives
\begin{equation}
\left[K_+,K_-\right]=\frac{iR}{\hbar } \equiv \frac{i}{L^2}\ ,
\label{DP1}
\end{equation}
and a canonical set of conjugate position coordinates
\begin{math} (\xi_+,\xi_-) \end{math} may be defined by
\begin{eqnarray}
\xi_\pm &=& \mp L^2K_\mp \nonumber \\
\left[\xi_+,\xi_-\right] &=& iL^2. \label{DP2}
\end{eqnarray}
Another independent canonical set of conjugate position
coordinates
\begin{math} (X_+,X_-) \end{math} is defined by
\begin{eqnarray}
x_+=X_+ +\xi_+ &,& x_-=X_- +\xi_-
\nonumber \\
\left[X_+,X_-\right] &=& -iL^2. \label{DP3}
\end{eqnarray}
Note that \begin{math} [X_a,\xi_b]=0 \end{math} , where
\begin{math} a=\pm \end{math} and  \begin{math} b=\pm \end{math}.

\begin{figure}[htbp]
\begin{center}
\mbox{\epsfig{file=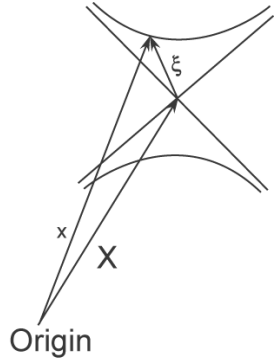,width=3.5cm}} \caption{The
hyperbolic path of a particle moving in the $x=(x_+,x_-)$ plane.
The noncommuting coordinate pairs  $X=(X_+,X_-)$, which points
from the origin to hyperbolic center, and $\xi =(\xi_+, \xi_- )$,
which points from the center of the orbit to the position on the
hyperbola, are shown. $x=X+\xi$.}
\label{ncfig2}
\end{center}
\end{figure}

The commutation relations Eqs.(\ref{DP2}) and (\ref{DP3})
characterize the noncommutative geometry in the plane $(x_+,x_-)$.
It is interesting to consider the case of pure friction in which
the potential
\begin{math} U=0 \end{math}. Eqs.(\ref{(7ba)}),
(\ref{momentum}) and (\ref{DP2}) then imply
\begin{equation}
{\cal H}_{friction}=\frac{\hbar^2}{2M}(K_+^2-K_-^2)
=-\frac{\hbar^2}{2ML^4}(\xi_+^2-\xi_-^2). \label{DP4}
\end{equation}
The equations of motion are
\begin{equation}
\dot{\xi}_\pm =\frac{i}{\hbar}\left[{\cal
H}_{friction},\xi_\pm\right] =-\frac{\hbar}{ML^2}\xi_\mp
=-\frac{R}{M}\xi_\mp =-\Gamma \xi_\mp , \label{DP5}
\end{equation}
with the solution
\begin{equation}
\left(\begin{array}{c}
  \xi_+(t)\\
  \xi_-(t) \\
\end{array}\right)=
\left(\begin{array}{cc}
  \cosh(\Gamma t) & -\sinh(\Gamma t) \\
  -\sinh(\Gamma t) & \cosh(\Gamma t) \\
\end{array}\right)
\left(\begin{array}{c}
  \xi_+ \\
  \xi_- \\
\end{array}\right).
\label{DP6}
\end{equation}
Eq.(\ref{DP6}) describes the hyperbolic orbit
\begin{equation}
\xi_-(t)^2-\xi_+(t)^2= \frac{2L^2}{\hbar \Gamma
}{\cal{H}}_{friction} \ . \label{DP7}
\end{equation}

The hyperbolae are defined by
\begin{math}
(x-X)^2-c^2(t-T)^{2}=\Lambda^2 , \label{PairProduct}
\end{math}
where \begin{math} \Lambda^2 = (\frac{mc}{\hbar} L^2)^2
\end{math}, the hyperbolic center is at \begin{math} (X,cT)
\end{math} and one branch of the hyperbolae is a particle
moving forward in time while the other branch is the same particle
moving backward in time as an anti-particle.

Now I observe that a quantum phase interference  of the
Aharanov-Bohm type can always be associated with the
noncommutative plane where
\begin{equation} \lab{1noncomm}
[X,Y]=iL^2~,
\end{equation}
with $L$ denoting the geometric length scale in the plane. Suppose
that a particle can move from an initial point in the plane to a
final point in the plane via one of two paths, say
\begin{math} {\cal P}_1  \end{math} or \begin{math} {\cal P}_2 \end{math}.
Since the paths start and finish at the same point, if one
transverses the first path in a forward direction and the second
path in a backward direction, then the resulting closed path
encloses an area \begin{math} {\cal A} \end{math}. The phase
interference $\vartheta$ between these two points is determined by
the difference between the actions for these two paths
\begin{math} \hbar \vartheta ={\cal S}({\cal P}_1)-{\cal S}({\cal P}_2)
\end{math}, and I show below it may be written as
\begin{equation}
\vartheta = \frac{\cal A}{L^2}~. \label{IntPhase}
\end{equation}

A physical realization of the mathematical noncommutative plane is
present in every laboratory wherein a charged particle moves in a
plane with a normal uniform magnetic field \begin{math} {\bf B}
\end{math}.  For this case, there are two canonical pairs of
position coordinates which do not commute: (i) the position
\begin{math} {\bf R} \end{math} of the center of the cyclotron
circular orbit and (ii) the radius vector \begin{math} {\bf \rho }
\end{math} from the center of the circle to the charged particle
position
\begin{math} {\bf r}={\bf R}+{\bf \rho } \end{math} (Fig.3).
The magnetic length scale of the noncommuting geometric
coordinates is due to Landau \cite{landau},
\begin{equation}
L^2=\frac{\hbar c}{eB}=\frac{\phi_0}{2\pi B} \ \ \ {\rm
(magnetic)}. \label{MagneticLength}
\end{equation}
Here \begin{math} \phi_0 \end{math} is the magnitude of the
magnetic flux quantum associated with a charge \begin{math} e
\end{math}.

\begin{figure}[htbp]
\begin{center}
\mbox{\epsfig{file=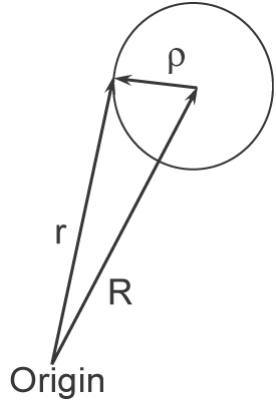,width=3.5cm}} \caption{A charge $e$
moving in a circular cyclotron orbit. Noncommuting coordinate
pairs are ${\bf R}=(X,Y)$, which points from the origin to the
orbit center, and ${\bf \rho }=(\rho_x,\rho_y)$, which points from
the center of the orbit to the charge position ${\bf r}={\bf
R}+{\bf \rho}$.} \label{ncfig1}
\end{center}
\end{figure}

For motion at fixed energy one may (in classical mechanics)
associate with each path
\begin{math} {\cal P} \end{math} (in phase space) a phase space action integral
\begin{equation}
{\cal S}({\cal P})=\int_{\cal P} p_i dq^i. \label{phase1}
\end{equation}

As said, the phase interference \begin{math} \vartheta
\end{math} between the two paths \begin{math} {\cal P}_1
\end{math} and \begin{math} {\cal P}_2 \end{math} is
determined by the action difference
\begin{equation}
\hbar \vartheta =\int_{{\cal P}_1} p_i dq^i-\int_{{\cal P}_2} p_i
dq^i =\oint_{{\cal P}=\partial \Omega } p_i dq^i \label{phase2}
\end{equation}
wherein \begin{math} {\cal P} \end{math} is the closed path which
goes from the initial point to the final point via path
\begin{math} {\cal P}_1 \end{math} and returns back to the initial point
via \begin{math} {\cal P}_2 \end{math}. The closed
\begin{math} {\cal P} \end{math} path may be regarded
as the boundary of a two-dimensional surface
\begin{math} \Omega \end{math}; i.e.
\begin{math} {\cal P}=\partial \Omega \end{math}.
Stokes theorem yields
\begin{equation}
\vartheta = \frac{1}{\hbar }\oint_{{\cal P}=\partial \Omega } p_i
dq^i =\frac{1}{\hbar }\int _\Omega (dp_i \wedge dq^i).
\label{phase3}
\end{equation}
The quantum phase interference \begin{math} \vartheta  \end{math}
between two alternative paths is thereby proportional to an
``area'' ${\cal A}$ of a surface \begin{math} \Omega  \end{math}
in phase space \begin{math} (p_1,\ldots ,p_f;q^1,\ldots ,q^f)
\end{math} as described by the right hand side of
Eq.(\ref{phase3}).

If one reverts to the operator formalism and writes the
commutation Eq.(\ref{1noncomm}) in the noncommutative plane as
\begin{equation}
[X,P_X]=i\hbar \ \ \ {\rm where} \ \ \ P_X=\left(\frac{\hbar
Y}{L^2}\right), \label{phase4}
\end{equation}
then back in the path integral formalism Eq.(\ref{phase3}) reads
\begin{equation}
\vartheta =\frac{1}{\hbar }\int _\Omega (dP_X \wedge dX)
=\frac{1}{L^2}\int _\Omega (dY \wedge dX)
\end{equation}
and Eq.(\ref{IntPhase}) is proved, i.e.  the quantum phase
interference between two alternative paths in the plane is
determined by the noncommutative length scale
\begin{math} L  \end{math} and the enclosed area
\begin{math} {\cal A} \end{math}.

I also remark that the existence of a phase interference  is
intimately connected to the zero point fluctuations in the
coordinates; e.g. Eq.(\ref{1noncomm}) implies a zero point
uncertainty relation
\begin{math} \Delta X \Delta Y \ge (L^2/2) \end{math}.

Resorting back to Eq.(\ref{DP1}) for the quantum dissipative case,
i.e.
\begin{equation}
L^2=\frac{\hbar}{R} \ \ \ {\rm (dissipative)}.
\label{DissipationLength}
\end{equation}
one then concludes that, provided $x_+ \neq x_-$, the quantum
dissipative phase interference   $\vartheta = \frac{\cal A}{L^2}
= \frac{{\cal A}R}{\hbar}$  is associated with two paths in the
noncommutative plane, starting at the same point ${\cal P}_1$ and
ending to the same point ${\cal P}_2$ so to enclose the surface of
area ${\cal A}$.

A comparison can be made between the noncommutative dissipative
plane and the noncommutative Landau magnetic plane as shown in
Fig.3. The circular orbit in Fig.3 for the magnetic problem is
replaced by the hyperbolic orbit and it may be shown that the
magnetic field is replaced by the electric field. The hyperbolic
orbit in Fig.2 is reflected in the classical orbit for a charged
particle moving along the \begin{math} x \end{math}-axis in a
uniform electric field. For more details on this comparison see
\cite{noncomm}.

Finally, I recall that the Lagrangian for the system of Eqs.
(\ref{QD12}) has been found \cite{Blasone:1996yh} to be the same
as the Lagrangian for three-dimensional topological massive
Chern-Simons gauge theory in the infrared limit. It is also the
same as for a Bloch electron in a solid which propagates along a
lattice plane with a hyperbolic energy surface
\cite{Blasone:1996yh}. In the Chern-Simons case one has
$\theta_{CS} = R/M = (\hbar/M L^2)$, with $\theta_{CS} $ the
``topological mass parameter''. In the Bloch electron case,
$(eB/\hbar c) =(1/L^2)$, with $B$ denoting the $z$-component of
the applied external magnetic field. In ref. \cite{Blasone:1996yh}
it has been considered the symplectic structure for the system of
Eqs. (\ref{QD12}) in the case of strong damping $R \gg M$ (the
so-called reduced case) in the Dirac constraint formalism as well
as in the Faddeev and Jackiw formalism \cite{Faddeev} and in both
formalisms a non-zero Poisson bracket for the ($x_+,~x_-$)
coordinates has been found.

Below I will consider the algebraic structure of the space of the
physical states emergent from the doubling of the degrees of
freedom discussed in the present and in the previous Section.
Before that I will discuss thermal field theory in the following
Section.

\section{Thermal field theory}
\label{sec:6a}

In this section I discuss the doubling of the degrees of freedom
in connection with thermal field theory. Specifically, I will
comment on the formalism of thermo field dynamics (TFD)
\cite{Umezawa,TFD,Banf}. In Section 8 it will be shown that the
algebraic structure on which the TFD formalism is based is
naturally provided by the $q$--deformed Hopf algebras for bosons
and for fermions (usually called $h_q(1)$ and $h_q(1|1)$,
respectively).

The central point in the TFD formalism is the possibility to
express the statistical average $<{\cal A}>$ of an observable
${\cal A}$ as the expectation value in the temperature dependent
vacuum $|0(\beta)>$:
\be <{\cal A}>~ \equiv~ {Tr[{\cal A}~ e^{-{\beta}{\cal H}}]
\over{Tr[e^{-{\beta}{\cal H}}]}}~ =~ <0(\beta)|{\cal A}|0(\beta)>
\; , \lab{p361} \ee
where ${\cal H}= H - \mu N$, with $\mu$ the chemical potential.

The first problem is therefore to construct a suitable temperature
dependent state $|0(\beta)>$ which satisfies Eq. (\ref{p361}),
namely
\be \lab{p361a} <0(\beta)|{\cal A}|0(\beta)> ={1
\over{Tr[e^{-{\beta}{\cal H}}]}} \sum_{n} <n|{\cal A}|n>
e^{-{\beta} E_{n}}~ , \ee
for an arbitrary variable $\cal A$, with
\be \lab{p361b} {\cal H}|n> =E_{n}|n> ~,~~~ <n|m> = \de_{nm} ~.\ee
Such a state cannot be constructed as long as one remains in the
original Fock space $\{ |n> \}$. To see this, let me closely
follow \cite{TFD}. One can expand $|0(\beta)>$ in terms of $|n>$
as
\be \lab{p361c} |0(\beta)> = \sum_{n} f_{n}(\beta)|n> ~ . \ee
Then, use of this equation into (\ref{p361a}) gives
\be \lab{p361d} f^{*}_{n}(\beta)f_{m}(\beta) = {1
\over{Tr[e^{-{\beta}{\cal H}}]}} e^{-{\beta} E_{n}}\de_{nm} ~ ,\ee
which is impossible to be satisfied by c-number functions
$f_{n}(\beta)$. However, Eq. (\ref{p361d}) can be regarded as the
orthogonality condition in a Hilbert space in which the expansion
coefficient $f_{n}(\beta)$ is a vector. In order to realize such a
representation it is convenient to introduce a dynamical system
identical to the one under study, namely to double the given
system. The quantities associated with the doubled system are
denoted by the tilde in the usual notation of TFD \cite{TFD}. Thus
the tilde-system is characterized by the Hamiltonian $\tilde H$
and the states are denoted by $|\tilde n>$, with
\be \lab{p361e} {\tilde {\cal H}}|\tilde n> =E_{n}|\tilde n> ~,~~~
<\tilde n|\tilde m> = \de_{nm} ~.\ee
where $E_{n}$ is  the same as the one appearing in Eq.
(\ref{p361b}) by definition. It is also assumed that non-tilde and
tilde operators are commuting (anti-commuting) boson (fermion)
operators. One then considers the space spanned by the direct
product $|n> \otimes ~|\tilde m> \equiv |n,\tilde m>$. The matrix
element of a bose-like operator $\cal A$ is then
\be \lab{p361f} <\tilde m, n|{\cal A}|n',{\tilde m}'> = <n|{\cal
A}|n'> \de_{mm'} ~,\ee
and the one of the corresponding ${\tilde {\cal A}}$ is
\be \lab{p361ff} <\tilde m, n|{\tilde {\cal A}}|n',{\tilde m}'> =
<\tilde m|{\tilde {\cal A}}|{\tilde m}'> \de_{nn'} ~.\ee
In TFD it turns out to be convenient to identify
\be \lab{p361g} < m|{\cal A}|n> = <\tilde n|{\tilde {\cal
A}}^{\dag}|\tilde m>  ~.\ee

Eq. (\ref{p361d}) is satisfied if one defines
\be \lab{p361h} f_{n}(\beta) = {1 \over{\sqrt{Tr[e^{-{\beta}{\cal
H}}]}}} e^{-{{\beta} E_{n}}\over 2}|\tilde n>~ ,\ee
and Eq. (\ref{p361a}) is obtained by using the definition
(\ref{p361h}) in  $|0(\beta)>$ given by (\ref{p361c}):
\be \lab{p361ca} |0(\beta)> = {1 \over{\sqrt{Tr[e^{-{\beta}{\cal
H}}]}}} \sum_{n} e^{-{{\beta} E_{n}}\over 2}|n,\tilde n> ~ . \ee
The vectors $|n>$ and $|\tilde n>$ thus appear as a pair in
$|0(\beta)>$. I remark that the formal r\^ole of the ``doubled"
states $|\tilde n>$ is merely to pick up the diagonal matrix
elements of ${\cal A}$. In this connection, thinking of the r\^ole
of the environment, which  is able to reduce the system density
matrix to its diagonal form in the QM decoherence processes
\cite{zurek}, it is remarkable that the doubled degrees of freedom
in TFD are indeed susceptible of being interpreted as the
environment degrees of freedom, as better specified in the
following.

It is useful to consider, as an example, the case of the number
operator. Let ${\cal A} \equiv N = a^{\dag}a$. For definiteness I
consider the boson case. Then the statistical average of $N$ is
the Bose-Einstein distribution $f_{B}(\om)$, where $\om$ denotes
the mode energy, $H = \om a^{\dag}a$,
\be <N>~ \equiv~ {Tr[N~ e^{-{\beta} H}] \over{Tr[e^{-{\beta}
H}]}}~ =~ <0(\beta)| N|0(\beta)> = \frac{1}{e^{\beta \om} - 1} =
f_{B}(\om) \; . \lab{p361bose} \ee
One then can show \cite{TFD} that, by setting
\be u(\beta) \equiv~  \sqrt{1+f_{B}(\om) } , ~~~ v(\beta) \equiv~
\sqrt{f_{B}(\om) }~ , \lab{p361bose1} \ee
\be u^{2}(\beta) - v^{2}(\beta) = 1 ~ , \lab{p361bose2} \ee
so that
\be u(\beta) = \cosh \theta(\beta) ~, ~~~ v(\beta) = \sinh
\theta(\beta) ~ , \lab{p361bose3} \ee
and defining
\be {\cal G} \equiv -i (a^{\dag}{\tilde a}^{\dag} - a{\tilde a} )
~ , \lab{p361bose2} \ee
the state $|0(\beta)>$ is formally given (at finite volume) by
\be \lab{p361cab} |0(\beta)> = e^{i\theta(\beta){\cal G}}|0> = {1
\over{u(\beta)}} \exp \Big( \frac{v(\beta)}{u(\beta)}
\Big)a^{\dag}{\tilde a}^{\dag} |0>  ~.  \ee
It is clear that the state $|0(\beta)>$ is not annihilated by $a$
and $\ti a$.  However, it is annihilated by the ``new" set of
operators $a(\theta)$ and ${\ti a}(\theta)$,
\be a(\theta)|0(\beta)> = 0 = {\ti a}(\theta)|0(\beta)> ~,
\lab{p324a} \ee
with
\bea \nonumber a(\theta) &=& \exp(i{\theta}{\cal G})~ a~
\exp(-i{\theta}{\cal G}) =  a ~{\rm cosh} ~\theta - {\tilde
a}^{\dagger}
~{\rm sinh} ~\theta ~ , \\
{\tilde a}(\theta) &=& \exp(i{\theta}{\cal G}) ~{\tilde a}~
\exp(-i{\theta}{\cal G}) ~ = {\tilde a} ~{\rm cosh} ~\theta -
 a^{\dagger} ~{\rm sinh} ~\theta ~,
\lab{p324} \eea
 \be [ a(\theta) ,
a^{\dagger}(\theta) ] = 1 ~, ~~[ {\tilde a}(\theta) , {\tilde
a}^{\dagger}(\theta) ] = 1 ~.~~ \lab{p322} \ee
All other commutators are equal to zero and $a(\theta)$ and
${\tilde a}(\theta)$ commute among themselves. Eqs. (\ref{p324})
are nothing but the Bogoliubov transformations of the $(a, {\tilde
a})$ pair into a new set of creation, annihilation operators. I
will show in Section 8 that the Bogoliubov-transformed operators
$a(\theta)$ and ${\tilde a}(\theta)$ are linear combinations of
the deformed coproduct operators.

The state $|0(\beta)>$ is not the vacuum (zero energy eigenstate)
of $H$ and of $\ti H$. It is, however, the zero energy eigenstate
for the ``Hamiltonian" $\hat{H}$, $\hat{H}|0(\beta)> = 0$, with
\be \hat{H} \equiv H - {\ti H} = \om (a^{\dag}a - {\tilde
a}^{\dag}{\tilde a}).
 ~, \lab{p324ab} \ee
The state $|0(\beta)>$ is called the thermal vacuum.

I note that in the boson case $J_1 \equiv {1\over
2}(a^{\dagger}{\tilde a}^{\dagger}  + a{\tilde a})$ together with
$J_2 \equiv {1\over 2}{\cal G}$ and $J_{3} \equiv {1 \over 2} ( N
+ {\tilde N} + 1)$ close the algebra $su(1,1)$. Moreover, ${\delta
\over {\delta \theta}}(N( \theta) - {\tilde N}(\theta)) = 0$ ,
with $(N(\theta) - {\tilde N}(\theta)) \equiv
(a^{\dagger}(\theta)a(\theta) - {\tilde
a}^{\dagger}(\theta){\tilde a} (\theta))$, consistently with the
fact that ${1 \over 4}(N - {\tilde N})^{2}$ is the $su(1,1)$
Casimir operator.

In the fermion case  $ J_{1} \equiv {1\over 2}{\cal G}$, $J_{2}
\equiv {1\over 2}(a^{\dagger}{\tilde a}^{\dagger}  + a{\tilde a})$
and $J_{3} \equiv {1 \over 2}( N  + {\tilde N} - 1)$ close the
algebra $su(2)$. Also in this case ${\delta \over
{\delta\theta}}(N(\theta) - {\tilde N}(\theta)) = 0$ , with
$(N(\theta) - {\tilde N}(\theta)) \equiv (a^{\dagger}(\theta)a(
\theta ) - {\tilde a}^{\dagger}(\theta){\tilde a}(\theta))$, again
consistently with the fact that ${1 \over 4}(N - {\tilde N})^{2}$
is related to the $su(2)$ Casimir operator.

Summarizing, the vacuum state for $a(\theta)$ and ${\tilde
a}(\theta)$ is formally given (at finite volume) by
\be |0(\theta)> ~=~ \exp ~(i{\theta}{\cal G})~ ~|0, 0> ~=~
\sum_{n} c_{n} (\theta )\, |n,{\tilde n}> ~, \lab{41} \ee
with $n, {\tilde n}= 0,..,\infty$ for bosons and $n, {\tilde n} =
0,1$ for fermions, and it appears therefore to be an $SU(1,1)$ or
$SU(2)$ generalized coherent state \cite{perelomov}, respectively
for bosons or for fermions.

In the infinite volume limit $|0(\theta)>$ becomes orthogonal to
$|0,0>$ and we have that the whole Hilbert space
$\{|0(\theta)>\}$, constructed by operating on $|0(\theta)>$ with
$a^{\dagger}(\theta)$ and ${\tilde a}^{\dagger}(\theta)$, is
asymptotically (i.e. in the infinite volume limit) orthogonal to
the space generated over $\{|0,0>\}$. In general, for each value
of $\theta (\beta)$, i.e. for each value of the  temperature  $T =
\frac{1}{k_{B}\beta}$, one obtains in the infinite volume limit a
representation of the canonical commutation relations unitarily
inequivalent to the others, associated with different values of
$T$. In other words, the parameter $\theta (\beta)$ (or the
temperature $T$) acts as a label for the inequivalent
representations \cite{Celeghini:1992yv}.

The TFD formalism is a fully developed QFT formalism
\cite{Umezawa,TFD,Banf} and it has been applied to a rich set of
problems of physical interest, in condensed matter physics, high
energy physics, quantum optics, etc. (see
\cite{Umezawa,TFD,Alfinito:2000bv,
Martellini:1978sm,Celeghini:1992yv,Celeghini:1989qc,Vitiello:1995wv,Banf}
and references therein quoted). I will show in Section 8 that the
doubling of the degrees of freedom on which the TFD formalism is
based finds its natural realization in the coproduct map.

Let me recall the so-called $''$tilde-conjugation rules$\, ''$
which are defined in TFD. For any two bosonic (respectively,
fermionic) operators ${\cal O}$ and ${\cal O'}$ and any two
$c$-numbers ${\alpha}$ and ${\beta}$ the tilde-conjugation rules
of TFD are postulated to be the following~\cite{TFD}:
\be ({\cal O}{\cal O'})^{\tilde {}} = {\tilde {{\cal O}}}{\tilde
{{\cal O'}}}~, \lab{p31} \ee
\be ({\alpha}{\cal O} +{\beta}{\cal O'})^{\tilde {}} =
{\alpha}^{*}{\tilde {{\cal O}}} + {\beta}^{*}{\tilde {{\cal
O'}}}~, \lab{p32} \ee
\be ({\cal O}^{\dagger})^{\tilde {}} = {\tilde {{\cal
O}}}^{\dagger}~, \lab{p33} \ee
\be ({\tilde {\cal O}})^{\tilde {}} = {\cal O}~. \lab{p34} \ee
According to (\ref{p31}) the tilde-conjugation does not  change
the order among operators. Furthermore, it is required that tilde
and non-tilde operators are mutually commuting (or anti-commuting)
operators and that the thermal vacuum $|0(\beta)>$ is invariant
under tilde-conjugation:
\be [ {\cal O} , {\tilde {\cal O'}}]_{\mp} = 0 = [ {\cal O} ,
{\tilde {\cal O'}}^{\dagger} ]_{\mp} ~, \lab{p35} \ee \be
{|0(\beta)>}^{\tilde {}} = |0(\beta)> ~. \lab{p36} \ee
In order to use a compact notation it is useful to introduce the
label $\sigma$ defined by $\sqrt{\sigma} \equiv +1$ for bosons and
$\sqrt{\sigma} \equiv + i$ for fermions. I shall therefore simply
write commutators as $[ {\cal O},{\cal O'} ]_{-\sigma} \doteq
{\cal O} {\cal O'}-\sigma {\cal O'}{\cal O}$, and $(1 \otimes
{\cal O} )({\cal O'} \otimes 1) \equiv\sigma ({\cal O'} \otimes
1)(1 \otimes {\cal O})$, without further specification of whether
${\cal O}$ and ${\cal O'}$ (which are equal to $a$, $a^{\dagger}$
in all possible ways) are fermions or bosons.

Upon identifying from now on $a_{1} \equiv a$, $a_{1}^{\dagger}
\equiv a^{\dagger}$, one easily checks that the TFD
tilde-operators (consistent with (\ref{p31}) -- (\ref{p36})) are
straightforwardly recovered by setting $a_{2} \equiv {\tilde a}$ ,
$a_{2}^{\dagger} \equiv {\tilde a}^{\dagger}$. In other words,
according to such identification, it is the action of the ~$1
\leftrightarrow 2$~ permutation $\pi$: $\pi a_{i} = a_{j} ,~~
i\neq j ,~~ i,j =1,2$, that realizes the operation of
$''$tilde-conjugation$\, ''$ defined in (\ref{p31} - \ref{p34}):
\be \pi a_{1} = \pi (a \otimes {\bf 1} ) = {\bf 1} \otimes a =
a_{2} \equiv {\tilde a} \equiv  {( a )^{\tilde {} }} \lab{p37} \ee
\be \pi a_{2} = \pi ({\bf 1} \otimes a) = a \otimes {\bf 1} =
a_{1} \equiv a \equiv {({\tilde a})^{\tilde {} }}  ~ . \lab{p38}
\ee

In particular, since the permutation $\pi$ is involutive, also
tilde-conjugation turns out to be involutive, as in fact required
by the rule (\ref{p34}). Notice that, as $(\pi a_{i})^{\dagger} =
\pi ({a_i}^{\dagger})$, it is also ${ ( { (a_i)^{\tilde {}} }~
)^{\dagger} } = ((a_i)^{\dagger})^{\tilde {} }$, i.e.
tilde-conjugation commutes with hermitian conjugation.
Furthermore, from  (\ref{p37})-(\ref{p38}), one has
\be {(ab)^{\tilde {} }} = {[(a \otimes {\bf 1})(b \otimes {\bf
1})]^{\tilde {}}} = {(ab \otimes {\bf 1})^{\tilde {} }} = {\bf 1}
\otimes ab = ({\bf 1} \otimes a)({\bf 1} \otimes b) = {\tilde
a}{\tilde b} \; . \lab{p39} \ee
Rules (\ref{p33}) and (\ref{p31}) are thus obtained. (\ref{p35})
is insured by the $\sigma$-commutativity of $a_{1}$ and $a_{2}$.
The vacuum of TFD,~ $|0(\beta)>$, ~is a condensed state of equal
number of tilde and non-tilde particles~\cite{TFD}, thus
(\ref{p36}) requires no further conditions: Eqs.
(\ref{p37})-(\ref{p38}) are sufficient to show that the rule
(\ref{p36}) is satisfied.

TFD appears equipped with a set of canonically conjugate
$''$thermal$\, ''$ variables: ${\theta}$ and
$\displaystyle{p_{\theta} \equiv -i{\delta \over {\delta
\theta}}}$~. $p_{\theta}$  can be regarded as the momentum
operator $''$conjugate$\, ''$ to the $''$thermal degree of
freedom$\, ''$ $\theta$. The notion of thermal degree of freedom
\cite{Banf} thus acquires formal definiteness in the sense of the
canonical formalism. It is remarkable that the "conjugate thermal
momentum" $p_{\theta}$ generates transitions among inequivalent
(in the infinite volume limit) representations: $\exp (i{\bar
\theta}p_{\theta})~ ~|0(\theta)> = |0(\theta + {\bar \theta})>$.
Notice that derivative with respect to the $\theta$ parameter is
actually a derivative with respect to the system temperature $T$.
This sheds some light on the r\^ole of $\theta$ in thermal field
theories for non-equilibrium systems and phase transitions. I
shall comment more on this point in the following Section.

Finally, when the proper field description is taken into account,
$a$ and ${\tilde a}$ carry dependence on the momentum ${\bf k}$.
The Bogoliubov transformation analogously, should be thought of as
inner automorphism of the algebra $su(1,1)_{\bf k}$ (or
$su(2)_{\bf k}$). This shows that one is globally dealing with
$\displaystyle{\bigoplus_{\bf k} su(1,1)_{\bf k}}$ (or
$\displaystyle{ \bigoplus_{\bf k} su(2)_{\bf k}}$). Therefore one
is lead to consider ${\bf k}$-dependence also for the ${\theta}$
parameter.

As a final comment, I observe that the "analogies" with the
formalism of quantum dissipation presented in Section 3.1 are
evident.

\section{The $q$--deformed Hopf algebra and QFT}
\label{sec:6}

In this Section I want to point out that the doubling of the
degrees of freedom is intimately related to the structure of the
space of the states in QFT \cite{Celeghini:1998sy}. This brings us
to consider the $q$--deformed Hopf algebra
\cite{25.,Celeghini:1991km}.

One key ingredient of Hopf algebra \cite{Celeghini:1991km} is the
coproduct operation, i.e. the operator doubling implied by the
coalgebra. The coproduct operation is indeed a map ${\Delta}:
{\cal A}\to {\cal A}\otimes {\cal A}$ which duplicates the algebra
${\cal A}$. Coproducts are commonly used in the familiar addition
of energy, momentum, angular momentum and of other so-called
primitive operators. The coproduct of a generic operator ${\cal
O}$ is a homomorphism defined as $\Delta {\cal O} = {\cal O}
\otimes {\bf 1} + {\bf 1} \otimes {\cal O} \equiv {\cal O}_1 +
{\cal O}_2$, with ${\cal O} \in {\cal A}$. Since additivity of
observables such as energy, momentum, angular momentum, etc. is an
essential requirement, the coproduct, and therefore the Lie-Hopf
algebra structure, appears to provide an essential algebraic tool
in QM and in QFT.

The systems discussed in the Sections above, where the duplication
of the degrees of freedom has revealed to be central, are thus
natural candidates to be described by the Lie-Hopf algebra. The
remarkable result holds \cite{Celeghini:1998sy} according to which
the infinitely many ui representations of the ccr, whose existence
characterizes QFT, are classified by use of the {\it
$q$--deformed} Hopf algebra. Quantum deformations of Hopf algebra
have thus a deeply non-trivial physical meaning in QFT.

In the following I consider boson operators. The discussion and
the conclusions can be easily extended to the case of fermion
operators \cite{Celeghini:1998sy}. For notational simplicity I
will omit the momentum suffix $\kappa$.

The bosonic algebra $h(1)$ is generated by the set of operators
$\{ a, a^{\dagger},H,N \}$ with commutation relations:
\be [\, a\, ,\, a^{\dagger} \, ] = \ 2H \, , \quad\; [\, \ N \,
,\, a \, ] = - a \, , \quad\; [\, \ N \, ,\, a^{\dagger} \, ] =
a^{\dagger} \, , \quad\; [\, \ H \, ,\, \bullet \, ] = 0 \, .
\lab{p22}
\ee
$H$ is a central operator, constant in each representation. The
Casimir operator is given by ${\cal C} = 2NH -a^{\dagger}a$.
~$h(1)$ is an Hopf algebra and is therefore equipped with the
coproduct operation, defined by
\be \Delta a = a \otimes {\bf 1} + {\bf 1} \otimes a \equiv a_1 +
a_2 ~,~~~ ~\Delta a^{\dagger} = a^{\dagger} \otimes {\bf 1} + {\bf
1} \otimes a^{\dagger} \equiv a_1^{\dagger} + a_2^{\dagger} ~,
\lab{p23}\ee \be \Delta H = H \otimes {\bf 1} + {\bf 1} \otimes H
\equiv H_{1} +
 H_{2}~, ~~~\Delta N = N \otimes {\bf 1} + {\bf 1} \otimes N \equiv  N_{1} +
 N_{2} ~.  \lab{p24}
 \ee
Note that $[a_i , a_j ] = [a_i , a_{j}^{\dagger} ] = 0 , ~i,j =
1,2,~ i \neq j $. The coproduct  provides the prescription for
operating on two modes.  As mentioned, one familiar example of
coproduct is the  addition of the angular momentum $J^{\alpha}$,
${\alpha} = 1,2,3$, of two particles: $\Delta J^{\alpha} =
J^{\alpha}  \otimes {\bf 1} + {\bf 1} \otimes J^{\alpha} \equiv
J^{\alpha}_1 + J^{\alpha}_2,~ J^{\alpha} \in su(2)$.

The $q$-deformation of $h(1)$ is the Hopf algebra $h_{q}(1)$:
\be [\, a_{q}\, ,\, a_{q}^\dagger \, ] = \ [2H]_{q} \, , \quad\;
[\, \ N \, \, , \, a_{q}\, ] = - a_{q} \, , \quad\;  [ \, \ N \, ,
\, a_{q}^\dagger \,] = a_{q}^\dagger , \quad\; [\, \ H \, \, , \,
\bullet \, ] = 0  \, , \lab{p26}
\ee
where $N_{q} \equiv N$ and $H_{q} \equiv H$.  The Casimir operator
${\cal C}_{q}$ is given by ${\cal C}_{q} = N[2H]_{q}
-a_{q}^{\dagger}a_{q}$, where $\displaystyle{[x]_{q} = {{q^{x} -
q^{-x}} \over {q - q^{-1}}}}$. The deformed coproduct is defined
by
\be \Delta a_{q} = a_{q} \otimes {q^{H}} + { q^{-H}} \otimes a_{q}
\, , \quad\quad \Delta a_{q}^{\dagger} = a_{q}^{\dagger} \otimes
{q^H} + {q^{-H}} \otimes a_{q}^{\dagger} ~, ~\lab{p28}
\ee
\be
\Delta H = H \otimes {\bf 1} + {\bf 1} \otimes H ~,~~~~ \Delta N =
N \otimes {\bf 1} + {\bf 1} \otimes N ~, \lab{p29}
\ee
whose algebra is isomorphic with (\ref{p26}): ~$ [ \Delta a_{q} ,
\Delta a_{q}^{\dagger} ] = [2 {\Delta} H]_{q}$ , etc. . Note that
$h_{q}(1)$ is a structure different from the commonly considered
$q$-deformation of the harmonic oscillator~\cite{B} that does not
have a coproduct and thus cannot allow for the duplication of the
state space.

I denote by ${\cal F}_{1}$  the single mode Fock space, i.e. the
fundamental representation $H = 1/2$, ${\cal C} = 0$. In such a
representation $h(1)$ and $h_{q}(1)$ coincide as it happens for
$su(2)$ and $su_{q}(2)$ for the spin-$\frac{1}{2}$ representation.
The differences appear in the coproduct and in the higher spin
representations.

As customary, I require that $a$ and $a^{\dag}$, and $a_{q}$ and
${a_{q}}^{\dag}$,  are adjoint operators. This implies that $q$
can only be real (or of modulus one in the fermionic case. In the
two mode Fock space ${\cal F}_{2} = {\cal F}_{1} \otimes {\cal
F}_{1}$, for $|q|=1$, the hermitian conjugation of the coproduct
must be supplemented by the inversion of the two spaces for
consistency with the coproduct isomorphism).

Summarizing, one can write for both bosons (and fermions) on
${\cal F}_{2} = {\cal F}_{1} \otimes {\cal F}_{1}$:
\be \Delta a =  a_1 + a_2 ~,~~~ ~\Delta a^{\dagger} =
a_1^{\dagger} + a_2^{\dagger} ~, \lab{p212} \ee \be \Delta a_{q} =
a_1 q^{1/2} + q^{-1/2} a_2 ~,~~~ ~\Delta a_{q}^{\dagger} =
a_1^{\dagger} q^{1/2}  +q^{-1/2} a_2^{\dagger} ~, \lab{p213} \ee
\be \Delta H = 1 , ~~~\Delta N = N_{1} +
 N_{2} ~.  \lab{pdelta} \lab{p214}
\ee
%

Now, the key point is \cite{Celeghini:1998sy} that the full set of
infinitely many unitarily inequivalent representations of the ccr
in  QFT are classified by use of the $q$--deformed Hopf algebra.
Since, as well known,  the Bogolubov transformations relate
different (i.e. unitary inequivalent) representations,  it is
sufficient to show that the Bogolubov transformations are directly
obtained by use of the deformed copodruct operation.  I consider
therefore the following operators (cf. (\ref{p28}) with $q(\theta)
\equiv e^{2\theta}$ and $H=1/2$):
\be {\alpha}_{q(\theta)} \equiv { { {\Delta} a_{q}} \over
{\sqrt{[2]_{q}} }} = {1\over\sqrt{[2]_{q}}} (e^{\theta} a_1 +
e^{-\theta} a_2 ) ~, \lab{p310} \ee
\be {\beta}_{q(\theta)} \equiv { 1 \over {\sqrt{[2]_{q}}} }
{\delta \over {\delta \theta}} {\Delta} a_{q} = {2q \over
\sqrt{[2]_{q}}}{{\delta}\over {\delta q}} \Delta a_{q} =
{1\over\sqrt{[2]_{q}}} (e^{\theta} a_1 -e^{-\theta} a_2 ) \; ,
\lab{p311} \ee
and h.c.. A set of commuting operators with canonical commutation
relations is given by
\be {\alpha}(\theta) \equiv {{\sqrt{[2]_{q}}}  \over 2{\sqrt2}} [
{\alpha}_{q(\theta)} + {\alpha}_{q(- \theta)} -
{\beta}_{q(\theta)}^{\dagger}  +  {\beta}_{q(- \theta)}^{\dagger}]
~, \lab{p314} \ee
\be {\beta}(\theta) \equiv {{\sqrt{[2]_{q}}} \over 2{\sqrt
2}}[{\beta}_{q(\theta)} + {\beta}_{q(- \theta)} -
{\alpha}_{q(\theta)}^{\dagger}  +  {\alpha}_{q(-
\theta)}^{\dagger} ] ~. \lab{p315} \ee
and h.c. One then introduces \cite{Celeghini:1998sy}
\bea A(\theta) \equiv \frac{1}{\sqrt{2}} \left ( {\alpha}(\theta )
+ {\beta}(\theta )\right ) &=& A ~{\rm cosh} ~\theta -
{B}^{\dagger}
~{\rm sinh} ~\theta ~~, \lab{p321a} \\
B(\theta) \equiv \frac{1}{\sqrt{2}} \left ( {\alpha}(\theta ) -
{\beta}(\theta ) \right ) &=& B ~{\rm cosh} ~\theta - A^{\dagger}
~{\rm sinh} ~\theta ~, \lab{p321} \eea
with
\be [ A(\theta) , A^{\dagger}(\theta) ] = 1 ~, ~~[ B(\theta)
, B^{\dagger}(\theta) ] = 1 ~.~~ \lab{p322} \ee
All other commutators are equal to zero and $A(\theta)$ and
$B(\theta)$ commute among themselves. Eqs. (\ref{p321a}) and
(\ref{p321})  are nothing but the Bogolubov transformations for
the $(A,B)$  pair (see the corresponding transformations, e.g. in
the case of the dho, Eqs. (\ref{(2.10a)}) and (\ref{(2.10b)})). In
other words, Eqs. (\ref{p321a}), (\ref{p321}) show that the
Bogolubov-transformed operators $A(\theta)$ and $B(\theta)$ are
linear combinations of the coproduct operators defined in terms of
the deformation parameter $q(\theta )$ and of their
${\theta}$-derivatives.

From this point on one can re-obtain the results discussed in the
previous Sections, for example for the dho provided one sets
$\theta \equiv \Gamma t$.

The generator of (\ref{p321a}) and (\ref{p321}) is ${\cal G}
\equiv -i(A^{\dagger}B^{\dagger} - AB)$:
\be - i{\delta \over {\delta \theta}} A(\theta) = [{\cal G},
A(\theta)] ~,~~~ - i{\delta \over {\delta \theta}} B(\theta) =
[{\cal G}, B(\theta)] ~, \lab{p326}\ee
and h.c.. Compare this generator with $H_{I}$ in Eq.
(\ref{(2.7)}).

Let $|0\rangle \equiv |0\rangle \otimes |0\rangle$ denote the
vacuum annihilated by $A$ and $B$, $A|0\rangle = 0 = B|0\rangle $.
By introducing the suffix $\kappa$ (till now omitted for
simplicity), at finite volume $V$ one obtains
\be\label{19}
 |0 (\theta) \rangle \, = e^{i \sum_{\kappa}\theta_{\kappa}
 {\cal G_{\kappa}}} |0 \rangle \,
  =\prod_k\frac{1}{\cosh\theta_{k}}e^{{\tanh\theta_{k}
 A_k^{\dagger} {B}_{k}^{\dagger}}} |0\rangle ~,
\ee
to be compared with Eq. (\ref{(2.12)}). $\theta$ denotes the set
$\{\theta_{\kappa} = \frac{1}{2}\ln q_{\kappa}, \forall \kappa \}$
and $\langle 0(\theta)|0(\theta)\rangle = 1$. The underlying group
structure is $\bigotimes_{\kappa} SU(1,1)_{\kappa}$ and the vacuum
$|0 (\theta) \rangle$ is an $SU(1,1)$ generalized coherent state
\cite{perelomov}. The $q$--deformed Hopf algebra is thus
intrinsically related to coherence and to the vacuum structure in
QFT.

In the infinite volume limit, the number of degrees of freedom
becomes uncountable infinite, and thus one obtains \cite{Umezawa,
TFD,Celeghini:1992yv} $ \langle
0(\theta)|0(\theta^{\prime})\rangle\to 0$ ~ as $V\to\infty, \quad
\forall ~ \theta, \theta^{\prime}, ~ \theta\ne \theta^{\prime}$.
By denoting with ${\cal H}_\theta$ the Hilbert space with vacuum
$|0 (\theta) \rangle$, ${\cal H}_\theta \equiv \{ |0 (\theta)
\rangle \}$, this means that ${\cal H_{\theta }}$ and ${\cal
H}_{\theta '}$ become unitarily inequivalent. In this limit, the
``points" of the space ${\cal H} \equiv \{ {\cal H}_\theta, ~
\forall ~ \theta \}$  of the infinitely many ui representations of
the ccr are labelled by the deformation parameter $\theta$
${}$\cite{Celeghini:1998sy,Celeghini:1992yv}. The space ${\cal H}
\equiv \{ {\cal H}_\theta, ~ \forall ~ \theta \}$ is called the
space of the representations.

I note that $\displaystyle{p_{\theta} \equiv -i{\delta \over
{\delta \theta}}}$ can be regarded \cite{Celeghini:1998sy} as the
momentum operator ``conjugate" to the ``degree of freedom"
$\theta$. For an assigned fixed value $\bar{\theta}$, it is
\be e^{i{\bar{\theta}} p_{\theta}} A(\theta) =
e^{i{\bar{\theta}}{\cal G}} A(\theta) e^{-i{\bar{\theta}}{\cal G}}
= A( \theta + {\bar {\theta}} ) , \lab{p327}\ee
and similarly for $B(\theta)$.

It is interesting to consider the case of time--dependent
deformation parameter. This immediately relates to the dissipative
systems considered in the previous Sections. The Heisenberg
equation for $A(t,{\theta} (t))$ is
$$
-i{\dot A}(t,{\theta}(t)) = -i{\delta \over {\delta t}}
A(t,{\theta}(t)) -i{{\delta \theta} \over {\delta t}} {\delta
\over {\delta \theta}}A(t, {\theta}(t))=
$$
%
\be \left [ H , A(t,{\theta}(t)) \right ] + {{\delta \theta} \over
{\delta t}} [{\cal G}, A(t,{\theta}(t)) ] = \left [ H + Q ,
A(t,{\theta}(t)) \right ] , \lab{42} \ee
and $\displaystyle{Q \equiv {{\delta \theta} \over {\delta t}}
{\cal G}}$ plays the role of the heat--term in dissipative
systems. $H$ is the Hamiltonian responsible for the time variation
in the explicit time dependence of $A(t,{\theta}(t))$.  $H + Q$
can be therefore identified with the free
energy~\cite{Celeghini:1992yv}: variations in time of the
deformation parameter involve dissipation. In thermal theories and
in dissipative systems the doubled modes $B$ play the role of the
thermal bath or environment.

Summarizing, QFT is characterized by the existence of ui
representations of the ccr \cite{Bratteli} which are related among
themselves by the Bogoliubov transformations. These, as seen
above, are obtained as linear combinations of the deformed
coproduct maps which express the doubling of the degrees of
freedom. Therefore one may conclude that the intrinsic algebraic
structure of QFT (independent of the specificity of the system
under study) is  the one of the $q$--deformed Hopf algebra. The ui
representations existing in QFT are related and labelled by means
of such an algebraic structure.

It should be stressed that the coproduct map is also essential in
QM in order to deal with a many mode system (typically, with
identical particles). However, in QM all the representations of
the ccr are unitarily equivalent and therefore the Bogoliubov
transformations induce unitary transformations among the
representations, thus preserving their physical content. The
$q$--deformed Hopf algebra therefore does not have that physical
relevance in QM, which it has, on the contrary, in QFT. Here, the
representations of the ccr, related through Bogoliubov
representations, are unitarily {\it inequivalent} and therefore
physically inequivalent: they represent different physical phases
of the system corresponding to different boundary conditions, such
as, for example, the system temperature. Typical examples are the
superconducting and the normal phase, the ferromagnetic and the
non-magnetic (i.e. zero magnetization) phase, the crystal and the
gaseous phase, etc.. The physical meaning of the deformation
parameter $q$ in terms of which ui representations are labelled is
thus recognized.

When the above discussion is applied to non-equilibrium (e.g.
thermal and/or dissipative) field theories it appears that the
couple of  conjugate variables $\theta$ and $p_{\theta} \equiv
-i\frac{\partial}{\partial \theta }$, with $\theta = \theta
(\beta(t))$ ($\beta(t) = {1 \over k_B T(t)}$), related to the
$q$--deformation parameter, describe trajectories in the space
$\cal H$ of the representations. In \cite{Vitiello:2003me} it has
been shown that there is a symplectic structure associated to the
"degrees of freedom" $\theta$ and that the trajectories in the
$\cal H$ space may exhibit properties typical of chaotic
trajectories in classical nonlinear dynamics. I will discuss this
in the following. In the next Section I present further
characterizations of the vacuum structure of the ui
representations in QFT.

\section{Entropy as a measure of the entanglement}
\label{sec:7}

In Section 3 I have shown that the time evolution of the state $|0
(t) \rangle$ is actually controlled by the entropy variations (cf.
Eq. (\ref{(2.17)})). I will shortly comment on the entropy in this
Section from a more general point of view, also in connection with
entanglement of the $A-B$ modes, since it appears as a structural
aspect of QFT related with the existence of the ui representations
of the ccr.

The state $|0 (\theta) \rangle $ may be  written as:
\be |0 (\theta) \rangle \, = \exp{\left ( - {1\over{2}} S_{A}
\right )} |\,{\cal I}\rangle  \, = \exp{\left ( - {1\over{2}}
S_{B} \right )} |\,{\cal I}\rangle  ~, \ee
\be S_{A} \equiv - \sum_{\kappa} \Bigl \{ A_{\kappa}^{\dagger}
A_{\kappa} \ln \sinh^{2} {\theta}_{\kappa} - A_{\kappa}
A_{\kappa}^{\dagger} \ln \cosh^{2} {\theta}_{\kappa} \Bigr \} ~.
\ee
Here $ |\,{\cal I}\rangle \, \equiv \exp {\left( \sum_{\kappa}
A_{\kappa}^{\dagger} {B}_{\kappa}^{\dagger} \right)} |0\rangle $
and $S_{B}$ is given by an expression similar to $S_{A}$, with
${B}_{\kappa}$ and ${B}_{\kappa}^{\dagger}$ replacing $A_{\kappa}$
and $A_{\kappa}^{\dagger}$, respectively. I simply write $S$ for
either $S_{A}$ or $S_{B}$. I can also write
\cite{Umezawa,TFD,Celeghini:1992yv}:
\be\lab{M3}
  |0 (\theta) \rangle = \sum_{n=0}^{+\infty} \sqrt{W_n} \left( |n \rangle
  \otimes |{n} \rangle  \right) ~,
\ee
\be\label{M4}
  W_n = \prod_k
  \frac{\sinh^{2n_k}\theta_{k}}{\cosh^{2(n_k+1)}\theta_{k}}\,,
\ee
with $n$ denoting the set $\{ {n}_{\kappa} \}$ and with $ 0 < W_n
< 1$
 and $\sum_{n=0}^{+\infty} W_n = 1$. Then
\be\lab{M5}  \langle 0(\theta) |S_{A}|0(\theta) \rangle =
\sum_{n=0}^{+\infty} W_n ln W_n ~, \ee
which confirms that $S$ can be interpreted as the entropy operator
\cite{Umezawa,TFD,Celeghini:1992yv}.

The state $|0(\theta)\rangle$ in Eq. (\ref{19}) can be also
written as \bea \label{ent}
 &&|0 (\theta) \rangle \,
= \left ( \prod_k\;\frac{1}{\cosh\theta_{k}}
\right) \\
\nonumber &&\times \left(|0 \rangle \otimes |{0} \rangle + \sum_k
\tanh\theta_{k}
  \left( | A_k  \rangle \otimes |{B}_{k} \rangle
  \right)  + \dots \right),
\eea
which clearly  cannot be factorized into the product of two
single-mode states. There is thus entanglement between the modes
$A$ and $B$: $|0(\theta)\rangle$ is an entangled state. Eq.
(\ref{M3}) and (\ref{M5}) then show that $S$  provides a measure
of the degree of entanglement.

I remark that the entanglement is truly realized in the infinite
volume limit where
\be\lab{ort2} \langle 0 (\theta)| 0  \rangle = e^{-
\frac{V}{(2\pi)^{3}}\int d^{3} {\kappa} \ln \cosh \theta_{\kappa}}
\mapbelow{V \rightarrow \infty} 0 ~, \ee
provided $\int d^{3} {\kappa} \ln \cosh \theta_{\kappa}$ is  not
identically zero. The probability of having the component state
$|n \rangle
  \otimes |{n} \rangle$  in the state $|0 (\theta) \rangle$ is
$W_n$. Since $W_n$ is a decreasing monotonic function of $n$, the
contribution of the states $|n \rangle
  \otimes |{n} \rangle$ would be
suppressed for large $n$ at finite volume. In that case, the
transformation induced by the unitary operator $G^{-1}(\theta)
\equiv \exp(- i \sum_{\kappa}\theta_{\kappa}
 {\cal G_{\kappa}})$ could disentangle the $A$ and $B$
sectors. However, this is not the case in the infinite volume
limit, where the summation extends to an infinite number of
components and Eq. (\ref{ort2}) holds (in such a limit Eq.
(\ref{19}) is only a formal relation since $G^{-1}(\theta)$ does
not exist as a unitary operator)\cite{Iorio:2004bt}.

It is interesting to note that, although the mode $B$ is related
with quantum noise effects (cf. the discussion in Section 5),
nevertheless the $A-B$ entanglement is not affected by such noise
effects. The robustness of the entanglement is rooted in the fact
that, once the infinite volume limit is reached, there is no
unitary generator able to disentangle the $A-B$ coupling.

\section{Trajectories in the $\cal H$ space}
\label{sec:8}

In this Section I want to discuss the chaotic behavior, under
certain conditions, of the trajectories in the $\cal H$ space. Let
me start by recalling some of the features of the $SU(1,1)$ group
structure (see, e.g., \cite{perelomov}).

$SU(1,1)$ realized on ${\it C} \times {\it C}$ consists of all
unimodular $2 \times 2$ matrices leaving invariant the Hermitian
form $|z_{1}|^{2} - |z_{2}|^{2}$, $z_{i} \in {\it C}, i=1,2$. The
complex $z$ plane is foliated under the group action into three
orbits: $X_{+} = \{z:|z|<1 \}$, $X_{-} = \{z:|z|>1 \}$ and $X_{0}
= \{z:|z|=1 \}$.

The unit circle $X_{+} = \{\zeta : |\zeta|<1 \}$, $\zeta \equiv
e^{i \phi} \tanh \theta$, is isomorphic to the upper sheet of the
hyperboloid which is the set $\bf H$ of pseudo-Euclidean bounded
(unit norm) vectors ${\bf n}: {\bf n} \cdot {\bf n} = 1$. $\bf H$
is a K\"ahlerian manifold with metrics
\be ds^{2} = 4\frac{{\partial}^{2} F}{\partial \zeta
\partial {\bar{\zeta}}} d\zeta \cdot d \bar{\zeta}~,
\ee
and
\be \lab{KP} F \equiv -\ln(1 - {|\zeta|}^{2}) \ee
is the K\"ahlerian potential. The metrics is invariant under the
group action \cite{perelomov}.

The K\"ahlerian manifold $\bf H$ is known to have a symplectic
structure. It may be thus considered as the phase space for the
classical dynamics generated by the group action \cite{perelomov}.

The $SU(1,1)$ generalized coherent states are recognized to be
``points" in $\bf H$ and transitions among these points induced by
the group action are therefore classical trajectories
\cite{perelomov} in $\bf H$ (a similar situation occurs
\cite{perelomov} in the $SU(2)$ (fermion) case).

Summarizing, the space of the unitarily inequivalent
representations of the ccr, which is the space of the $SU(1,1)$
generalized coherent states, is a K\"ahlerian manifold, ${\cal H}
\equiv \{ {\cal H}_\theta, ~ \forall \theta \} \approx {\bf H}$;
it has a symplectic structure and a classical dynamics is
established on it by the $SU(1,1)$ action (generated by $\cal G$
or, equivalently, by $p_{\theta}$: ${\cal H}_{\theta} \rightarrow
{\cal H}_{\theta'}$). Variations of the $\theta$--parameter induce
transitions through the representations ${\cal H}_{\theta} = \{{|0
(\theta)\rangle}\}$, i.e. through the physical phases of the
system, the system order parameter being dependent on $\theta$.
These transitions are described as trajectories through the
``points" in ${\cal H}$. One may then assume time-dependent
$\theta$: $\theta = \theta (t)$. For example, this is the case of
dissipative systems and of non-equilibrium thermal field theories
where $\theta_{\kappa} = \theta_{\kappa} (\beta(t))$, with
$\beta(t) = {1\over k_B T(t)}$.

It is interesting to observe that, considering the transitions
${\cal H}_{\theta} \rightarrow {\cal H}_{\theta'}$, i.e. $|0 (
{\theta})\rangle \rightarrow |0 ({\theta'})\rangle$, we have
\be\lab{trans} \langle 0 (\theta)| 0 (\theta') \rangle =  e^{-
\frac{V}{2(2\pi)^{3}}\int d^{3} {\kappa}
F_{\kappa}(\theta,\theta')} \ee
where $F_{\kappa}(\theta,\theta')$ is given by Eq. (\ref{KP}) with
$|\zeta_{\kappa}|^{2} = \tanh^{2}(\theta_{\kappa} -
\theta'_{\kappa})$, which shows the role played by the K\"ahlerian
potential in the motion over $\cal H$.

The result that the group action induces classical trajectories in
$\cal H$ has been also obtained elsewhere \cite{Manka,DelGiudice}
on the ground of more phenomenological considerations.

With reference to the discussion presented in Sections 3 - 5, we
may say that on the (classical) trajectories in $\cal H$ it is
$x_+ = x_- = x_{classical}$, i.e. on these trajectories the
quantum noise accounted for by $y$ is fully shielded by the
thermal bath (cf. Eq. (\ref{(8)})). In Section 5 (see
\cite{Srivastava:1995yf}) it has been indeed observed that the $y$
freedom contributes to the imaginary part of the action which
becomes negligible in the classical regime, but is relevant for
the quantum dynamics, namely in each of  the ``points'' in $\cal
H$ (i.e. in each of the spaces ${\cal H}_{\theta}$, for each
$\theta$) through which the trajectory goes as $\theta$ changes.
Upon ``freezing'' the action of $G(\theta)$ (i.e. upon
``freezing'' the ``motion'' through the ui representations) the
quantum features of ${\cal H}_{\theta}$, at given ${\theta}$,
become manifest. This relates to the 't Hooft picture
\cite{'tHooft:1999bx} and to the results of refs.
\cite{Blasone:2000ew,Blasone:2002hq} where dissipation loss in
deterministic systems may manifest itself as quantum behavior (see
Section 11).

Let me use the notation $|0 (t)\rangle_{\theta} \equiv |0 (\theta
(t))\rangle$. For any $\theta(t) = \{\theta_{\kappa}(t), \forall
\kappa \}$ it is
\be\lab{nr} {}_{\theta}\langle 0(t) | 0(t)\rangle_{\theta}  = 1
~,~~\forall t~. \ee
I will now restrict the discussion to the case in which, for any
$\kappa$, $\theta_{\kappa} (t)$ is a growing function of time and
\be\lab{cond} \theta (t) \neq \theta (t')~, ~~\forall  t \neq
t',~~~ {\rm and} ~~~\theta (t) \neq \theta^{\prime} (t')
~,~~\forall t,t' ~. \ee
Under such conditions the trajectories in $\cal H$ satisfy the
requirements for chaotic behavior in classical nonlinear dynamics.
These requirements are the following \cite{hilborn}:

i)~~ the trajectories are bounded and each trajectory does not
intersect itself.

ii)~~trajectories specified by different initial conditions do not
intersect.

iii) trajectories of different initial conditions are diverging
trajectories.

Let $t_{0} = 0$ be the initial time. The "initial condition" of
the trajectory is then specified by the $\theta(0)$-set,
$\theta(0) = \{\theta_{\kappa}(0), \forall \kappa \}$. One obtains
\be\lab{tt} {}_{\theta}\langle 0(t) | 0(t') \rangle_{\theta}
 \mapbelow{V \rightarrow \infty} 0 ~, ~~ \forall \, t\, , t' ~ , ~~~
{\rm with}~~~t \neq t' ~ , \ee
provided $ {\int \! d^{3} \kappa \, \ln \cosh
(\theta_{\kappa}(t)-\theta_{\kappa}(t'))}$ is finite and positive
for any $t \neq t'$ .

Eq. (\ref{tt}) expresses the unitary inequivalence of the states
$|0(t)\rangle_{\theta} $ (and of the associated Hilbert spaces
$\{| 0(t) \rangle_{\theta} \}$) at different time values $t \neq
t'$ in the infinite volume limit. The non-unitarity of time
evolution, implied for example by the damping, is consistently
recovered in the unitary inequivalence among representations $\{|
0(t) \rangle_{\theta} \}$'s at different $t$'s in the infinite
volume limit.

The trajectories are  bounded in the sense of Eq. (\ref{nr}),
which shows that the ``length" (the norm) of the ``position
vectors" (the state vectors at time $t$) in $\cal H$ is finite
(and equal to one) for each $t$. Eq. (\ref{nr}) rests on the
invariance of the Hermitian form $|z_{1}|^{2} - |z_{2}|^{2}$,
$z_{i} \in {\it C}, i=1,2$ and I also recall that the manifold of
points representing the coherent states $|0(t)\rangle_{\theta}$
for any $t$ is isomorphic to the product of circles of radius
${r_{\kappa}}^{2} = \tanh^{2}(\theta_{\kappa}(t))$ for any
$\kappa$.

Eq. (\ref{tt}) expresses the fact that the trajectory does not
crosses itself as time evolves (it is not a periodic trajectory):
the ``points" $\{|0(t)\rangle_{\theta}\}$ and
$\{|0(t')\rangle_{\theta}\}$ through which the trajectory goes,
for any $t$ and $t'$, with $t \neq t'$, after the initial time
$t_{0} = 0$, never coincide. The requirement $i)$ is thus
satisfied.

In the infinite volume limit, we also have
\be\lab{ttn} {}_{\theta}\langle 0(t) | 0(t') \rangle_{\theta ' }
\mapbelow{V \rightarrow \infty} 0  \quad ~~ \forall \, t\, , t'
\quad , ~~\forall \, {\theta} \neq {\theta^{\prime}} ~. \ee
Under the assumption (\ref{cond}), Eq. (\ref{ttn}) is true also
for $t = t'$. The meaning of Eqs. (\ref{ttn}) is that trajectories
specified by different initial conditions ${\theta}(0) \neq
{\theta^{\prime}(0)}$ never cross each other. The requirement ii)
is thus satisfied.

In order to study how the ``distance" between trajectories in the
space $\cal H$ behaves as time evolves, consider two trajectories
of slightly different initial conditions, say ${{\theta}^{\prime}
(0)} = {\theta} (0) + \delta \theta$, with small $\delta \theta$.
A difference between the states $| 0(t)\rangle_{\theta}$ and $|
0(t)\rangle_{\theta^{\prime}}$ is the one between the respective
expectation values of the  number operator $A_{\kappa}^{\dagger}
A_{\kappa}$. For any $\kappa$ at any given $t$, it is
$$
\Delta {\cal N}_{A_{\kappa}}(t) \equiv {\cal
N'}_{A_{\kappa}}\bigl(\theta'(t)\bigr ) - {\cal
N}_{A_{\kappa}}\bigl(\theta(t)\bigr )
$$
\be = {}_{\theta^{\prime}}\langle 0(t) | A_{\kappa}^{\dagger}
A_{\kappa} |0(t)\rangle_{\theta^{\prime}} - {}_{\theta}\langle
0(t) |A_{\kappa}^{\dagger} A_{\kappa} | 0(t) \rangle_{\theta}
\lab{co1} \ee
\be \nonumber = \sinh^{2}\bigl ( {\theta^{\prime}}_{\kappa}(t)
\bigr ) - \sinh^{2}\bigl ( {\theta}_{\kappa}(t) \bigr ) = \sinh
\bigl ( 2{\theta}_{\kappa}(t)\bigr ){\delta \theta_{\kappa}(t)} ~,
\ee
where ${\delta \theta_{\kappa}(t)} \equiv
{\theta^{\prime}}_{\kappa}(t) - {\theta}_{\kappa}(t)$ is assumed
to be greater than zero, and the last equality holds for ``small"
${\delta \theta_{\kappa}(t)}$ for any $\kappa$ at any given $t$.
By assuming that $\frac{\partial{\delta \theta_{\kappa}}}{\partial
t}$ has negligible variations in time, the time-derivative gives
\be\lab{co3} \frac{\partial}{\partial t} \Delta {\cal
N}_{A_{\kappa}}(t) = 2 \frac{\partial
{\theta}_{\kappa}(t)}{\partial t }\cosh \bigl ( 2
{\theta}_{\kappa}(t)  \bigr ){\delta \theta_{\kappa}} ~. \ee
This shows that, provided $\theta_{\kappa}(t)$ is a growing
function of $t$, small variations in the initial conditions lead
to growing in time $\Delta {\cal N}_{A_{\kappa}}(t)$, namely to
diverging trajectories as time evolves.

In the assumed hypothesis, at enough large $t$ the divergence is
dominated by $\exp{(2\theta_{\kappa}(t)) }$. For each $\kappa$,
the quantity $2\theta_{\kappa}(t) $  could be thus thought to play
the r\^ole similar to the one of the Lyapunov exponent.

Since  $\sum_{\kappa} E_{\kappa} {\dot {\cal N}}_{A_{\kappa}} dt =
\frac{1}{\beta} dS_{A}$  , where $E_{\kappa}$ is the energy of the
mode $A_{\kappa}$ and $dS_{A}$ is the entropy variation associated
to the modes $A$ (cf. Eq. (\ref{dq})) \cite{Celeghini:1992yv}, the
divergence of trajectories of different initial conditions may be
expressed in terms of differences in the variations of the entropy
(cf. Eqs. (\ref{co1}) and (\ref{co3})):
\be\lab{co4} \Delta \sum_{\kappa} E_{\kappa}{\dot {\cal
N}}_{A_{\kappa}}(t) dt  = \frac{1}{\beta} \bigl ( dS'_{A} - dS_{A}
\bigr ) ~. \ee
The discussion above thus shows that also the requirement iii) is
satisfied. The conclusion is that trajectories in the $\cal H$
space exhibit, under the condition (\ref{cond}) and with
$\theta(t)$ a growing function of time, properties typical of the
chaotic behavior in classical nonlinear dynamics.

\section{Deterministic dissipative systems and quantization}

In Section 3 we have seen that the canonical quantization for the
damped oscillator is obtained at the expense of introducing an
``extra'' coordinate $y$. The role of the ``doubled" $y$
coordinate is absolutely crucial in the quantum regime where it
accounts for the quantum noise. When the classical solution $y =
0$ is adopted, the $x$ system appears to be ``incomplete"; the
loss of information due to dissipation  amounts to neglecting the
bath and to the ignorance  of the bath-system interaction, i.e.
the ignorance of ``where" and ``how" energy flows out of the
system. One can thus conclude that the loss of information
occurring at the classical level due to dissipation manifests
itself in terms of ``quantum" noise effects arising from the
imaginary part of the action, to which the $y$ contribution is
crucial. This result suggests to consider the approach to
dissipation presented above in connection with the  proposal put
forward by 't Hooft in a series of papers \cite{'tHooft:1999bx}.
He proposes that Quantum Mechanics may indeed result from a more
fundamental deterministic theory as an effect of a  process of
information loss. He considers a class of deterministic
Hamiltonian systems described by means of Hilbert space
techniques. The quantum systems are obtained when constraints
implementing the information loss are imposed on the original
Hilbert space. The Hamiltonian for such systems is of the form
\be \label{thofH} H = \sum_{i}p_{i}\, f_{i}(q)\,, \ee
where $f_{i}(q)$ are non--singular functions of the canonical
coordinates $q_{i}$. The equations for the $q$'s (i.e.
$\dot{q_{i}} = \{q_{i}, H\} = f_{i}(q)$) are decoupled from the
conjugate momenta $p_i$ and this implies \cite{'tHooft:1999bx}
that the system can be described deterministically even when
expressed in terms of operators acting on the Hilbert space. The
condition for the deterministic description is the existence of a
complete set of observables commuting at all times, called {\em
beables} \cite{Bell:1987hh}. For the systems of Eq.(\ref{thofH}),
such a set is given by the $q_i(t)$ \cite{'tHooft:1999bx}.

In order to cure the fact that the Hamiltonians of the type
(\ref{thofH}) are not bounded from below, one might split $H$ in
Eq.(\ref{thofH}) as \cite{'tHooft:1999bx}:
\bea && H = H_I - H_{II} \quad,\quad H_I = \frac{1}{4\rho}\left(
\rho + H \right)^{2}\;\; ,\;\;\; H_{II} = \frac{1}{4\rho}\left(
\rho - H\right)^{2}\, ,  \label{1.5} \eea
where  $\rho$ is a time--independent, positive function of
$q_{i}$.  $H_I$ and $H_{II}$ are then positively (semi)definite
and $ \{ H_I, H_{II} \} = \{\rho , H \} =  0$ . Then the
constraint condition is imposed onto the Hilbert space:
\be H_{II} |\psi \ran = 0\, , \label{3.8} \ee
which ensures that the Hamiltonian is bounded from below. This
condition, indeed, projects out the states responsible for the
negative part of the spectrum. In other words, one gets rid of the
unstable trajectories \cite{'tHooft:1999bx}.

In refs. \cite{Blasone:2000ew} and \cite{Blasone:2002hq} it has
been shown that the  system of damped-antidamped oscillators
discussed in Section 3 does provide  an explicit realization of 't
Hooft mechanism. In addition, it has been also shown that there is
a connection between the zero point energy of the quantum harmonic
oscillator and the geometric phase of the (deterministic) system
of damped/antidamped oscillators. This can be seen by noticing
that the Hamiltonian Eq.(\ref{(2.71)}) is of the type
(\ref{thofH}) with $i=1,2$ and with $f_1(q)=2\Om$, $f_2(q)=-2\Ga$,
provided one uses a set of canonical transformations which for
brevity I do not report here (see \cite{Blasone:2000ew}). By using
$J_2 = -\frac{i}{2}(J_{+} - J_{-})$ and ${\cal C} = {1\over{2}} (
A^{\dagger} A - B^{\dagger} B) $ one may write Eq. (\ref{(2.7)})
as
\be H = H_{I} - H_{II} \quad, \quad H_{I} = \frac{1}{2 \Om {\cal
C}} (2 \Om {\cal C} - \Ga J_2)^2\;\;\; ,\;\;\;\; H_{II} =
\frac{\Ga^2}{2 \Om {\cal C}} J_2^2\, .  \label{split} \ee
Note that ${\cal C}$, being the Casimir operator, is a constant of
motion, which ensures that once it has been chosen to be positive
it will remain such at all times. The constraint (\ref{3.8}) is
now imposed by putting
\be \label{thermalcondition}\quad J_2 |\psi\ran = 0\, , \ee
and the physical states $|\psi\ran$ are by this defined. It is now
convenient to introduce
\begin{eqnarray*}
x_1 = \frac{x+ y}{\sqrt{2}}\, , \,\,\,\,\ x_2 =
\frac{x-y}{\sqrt{2}}\, ,
\end{eqnarray*}
and
\be
 x_1 \ =\  r \cosh u \, , \,\,\,\,
x_2 \ = \ r \sinh u \,  , \ee
in terms of which \cite{Blasone:1996yh}
\bea \label{pu} {\cal C}  \ = \ \frac{1}{4 \Om m}\left[ p_r^2 -
\frac{1}{r^2}p_u^2 + m^2\Om^2 r^2\right]\, , \,\,\,\,  J_2 \ = \
\frac{1}{2}\, p_u\, . \eea
Of course, only  nonzero $r^{2}$  should be taken into account in
order for $\cal C$ to be invertible. If one does not use the
operatorial formalism, then the constraint $p_u=0$ implies $u =
-\frac{\ga}{2 m} t$. Eq.(\ref{thermalcondition}) implies
\be \lab{17} H |\psi\ran= H_{I} |\psi\ran=  2\Om {\cal C}|\psi\ran
= \left( \frac{1}{2m}p_{r}^{2} + \frac{K}{2}r^{2}\right) |\psi
\ran \, , \ee
where  $K\equiv m \Om^2$. $H_{I}$ thus reduces to the Hamiltonian
for the linear harmonic oscillator $\ddot{r} + \Om^2 r =0 $. The
physical states are even with respect to time-reversal
($|\psi(t)\ran  = |\psi(-t)\ran$) and periodical with period $\tau
= \frac{2\pi}{\Omega}$.

I will now introduce the states $|\psi(t)\ran_{H}$  and
$|\psi(t)\ran_{H_{I}}$ satisfying the equations:
\bea \lab{S1} i \hbar \frac{d}{dt} |\psi(t)\ran_{H} &=& H
\,|\psi(t)\ran_{H}~,
\\ \lab{S2}
i \hbar \frac{d}{dt} |\psi(t)\ran_{H_{I}} &= &2 \Om {\cal C}
|\psi(t)\ran_{H_{I}} \, . \eea
Eq.(\ref{S2}) describes the two-dimensional ``isotropic'' (or
``radial'') harmonic oscillator. $ H_{I} = 2 \Om{\cal C} $ has the
spectrum ${\cal H}^n_{I}= \hbar \Om n$, $n = 0, \pm 1, \pm 2,
...$. According to the choice for ${\cal C}$ to be positive, only
positive values of $n$ will be considered. The generic state
$|\psi(t)\ran_{H}$ can be written as
\bea\lab{eqt0} |\psi(t)\ran_{H} = {\hat{T}}\left[ \exp\left(
\frac{i}{\hbar}\int_{t_0}^t 2 \Ga J_2 dt' \ri) \ri]
|\psi(t)\ran_{H_{I}} ~, \eea
where ${\hat{T}}$ denotes time-ordering. Of course, here $\hbar$
is introduced on purely dimensional grounds and its actual value
cannot be fixed by the present analysis.

\begin{figure}[b]
\begin{center}
\includegraphics[width=.7\textwidth]{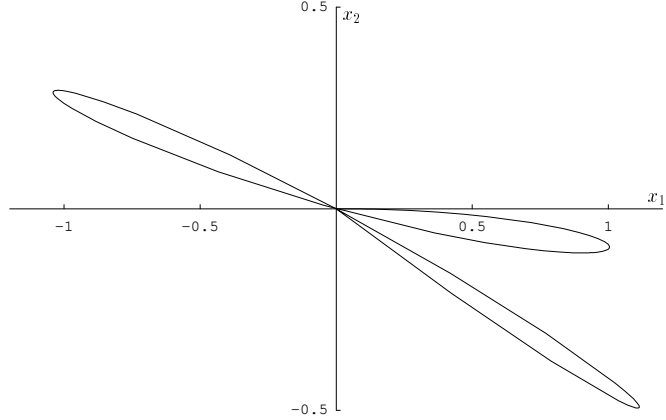}
\end{center}
\caption[]{Trajectories for $r_0=0$ and $v_0=\Om$, after three
half-periods for $\ka =20$, $\ga=1.2$ and $m=5$. The ratio
$\int_0^{\tau/2}({\dot x}_1 x_2 - {\dot x}_2 x_1)dt /{\cal E}
=\pi\frac{\Ga}{m\Om^3}$  is preserved. ${\cal E}$ is the initial
energy: ${\cal E}=\frac{1}{2} m v_0^2 + \frac{1}{2} m \Om^2
r_0^2$.} \label{eps1}
\end{figure}

One obtains \cite{Blasone:2000ew}:
\bea &&\,_{H}\lan \psi(\tau) | \psi(0) \ran_{H} =\,_{H_{I}} \lan
\psi(0)| \exp\left( i \int_{C_{0\tau}} A(t') dt' \right)
|\psi(0)\ran_{H_{I}} \equiv e^{i \phi} \, , \label{berryphase}
\eea
where the contour $C_{0\tau}$ is the one going from $t'=0$ to
$t'=\tau$ and back and  $A(t) \equiv \frac{\Ga m}{ \hbar}({\dot
x}_1 x_2 - {\dot x}_2 x_1)$. Note that $({\dot x}_1 x_2 - {\dot
x}_2 x_1) dt$ is the area element in the $(x_1,x_2)$ plane
enclosed by the trajectories (see Fig.4) (cf. Section 6). Notice
also that the evolution (or dynamical) part of the phase does not
enter in $\phi$, as the integral in Eq.(\ref{berryphase}) picks up
a purely geometric contribution \cite{Berry}.

Let me consider the periodic physical states $|\psi\rangle$.
Following \cite{Berry}, one writes
\bea |\psi(\tau)\ran &= &e^{ i\phi -
\frac{i}{\hbar}\int_{0}^{\tau}\langle \psi(t)| H |\psi(t) \rangle
dt} |\psi(0)\ran \,=\, e^{- i2\pi n} | \psi(0)\ran \, ,
\label{eqtH} \eea
i.e. $\frac{ \langle \psi(\tau)| H |\psi(\tau) \rangle  }{\hbar}
\tau - \phi = 2\pi n$, $n = 0,  1,  2, \ldots $,  which by using
$\tau = \frac{2 \pi}{\Om}$ and $\phi = \alpha \pi$, gives
\be \lab{spectrum} {\cal H}_{{_I},e\!f\!f}^n \equiv \langle
\psi_{n}(\tau)| H |\psi_{n}(\tau) \rangle= \hbar \Om \left( n +
\frac{\alpha}{2} \right) ~. \ee
The index $n$ has been introduced to exhibit the $n$ dependence of
the state  and  the corresponding energy. ${\cal
H}_{\I,e\!f\!f}^n$ gives the effective $n$th energy level of the
physical system, i.e. the energy given by ${\cal H}_{\I}^n$
corrected by its interaction with the environment. One thus see
that the dissipation term $J_2$ of the Hamiltonian  is actually
responsible for the ``zero point energy" ($n = 0$): $E_{0}
=\frac{\hbar}{2} \Om \alpha$.

I recall that the zero point energy is the ``signature" of
quantization since in Quantum Mechanics it is formally due to the
non-zero commutator of the canonically conjugate $q$ and $p$
operators. Thus dissipation manifests itself as ``quantization".
In other words, $E_0$, which appears as the ``quantum
contribution" to the spectrum, signals the underlying dissipative
dynamics.  If one wants to match the Quantum Mechanics zero point
energy, has to fix $\alpha = 1$, which gives \cite{Blasone:2000ew}
$\Om = \frac{\ga}{m}$.

In connection with the discussion presented in Section 3.1, the
thermodynamical features of the dynamical r\^ole of $J_2$ can be
revealed by rewriting Eq.(\ref{eqt0}) as
\bea\lab{eqt0U} |\psi(t)\ran_{H} = {\hat T}\left[ \exp\left(i
\frac{1}{\hbar} \int_{u(t_0)}^{u(t)} 2 J_2 du'\right) \right]
|\psi(t)\ran_{H_{I}} \, , \eea
where $u(t) = - \Ga t$ has been used. Thus,
\bea \lab{Su} -i \hbar \frac{\pa}{\pa u} |\psi(t)\ran_{H} = 2J_{2}
|\psi(t)\ran_{H} \, . \eea
$2 J_2$ appears then to be responsible for shifts (translations)
in the $u$ variable, as it has to be expected since $2 J_{2} =
p_{u}$ (cf. Eq.(\ref{pu})). One can write indeed: $p_{u} =- i
\hbar \frac{\pa}{\pa u}$. Then, in full generality,
Eq.(\ref{thermalcondition}) defines families of physical states,
representing stable, periodic trajectories (cf. Eq.(\ref{17})). $2
J_{2}$ implements transition from family to family, according to
Eq.(\ref{Su}). Eq.(\ref{S1}) can be then rewritten as
\bea \lab{S11} i \hbar \frac{d}{dt} |\psi(t)\ran_{H} = i \hbar
\frac{\pa}{\pa t} |\psi(t)\ran_{H} + i \hbar
\frac{du}{dt}\frac{\pa}{\pa u} |\psi(t)\ran_{H}\, . \eea
The first term on the r.h.s. denotes of course derivative with
respect to the explicit time dependence of the state. The
dissipation contribution to the energy is thus described by the
``translations" in the $u$ variable.  Now I consider the
derivative
\bea\lab{T} \frac{\pa S}{\pa U} = \frac{1}{T}\,. \eea
From Eq.(\ref{split}), by using $S \equiv \frac{2 J_{2}}{\hbar}$
and $U \equiv  2 \Om {\cal C}$, one obtains $T = \hbar \Ga$. Eq.
(\ref{T}) is the defining relation for temperature in
thermodynamics  (with $k_B = 1$) so that one could formally regard
$\hbar \Ga$ (which dimensionally is an energy) as the temperature,
provided the dimensionless quantity $S$ is identified with the
entropy. In such a case, the ``full Hamiltonian'' Eq.(\ref{split})
plays the role of the free energy ${\cal F}$: $H = 2 \Om {\cal C}
- (\hbar \Ga) \frac{2 J_2}{\hbar} = U - TS = {\cal F}$.  Thus $2
\Ga J_{2}$ represents the heat contribution in $H$ (or $\cal F$).
Of course, consistently, $\left. \frac{\pa {\cal F} }{\pa T
}\right|_\Om = - \frac{2 J_2}{\hbar}$.  In conclusion $\frac{2
J_{2}}{\hbar}$ behaves as the entropy, which is not surprising
since it controls the dissipative (thus irreversible) part of the
dynamics. In this way the conclusions of Section 3 are reobtained.
It is also suggestive that the temperature $\hbar \Ga$ is actually
given by the background zero point energy: $\hbar \Ga =
\frac{\hbar \Om}{2}$.

Finally, I observe that
\bea \left. \frac{\pa {\cal F}}{\pa \Om}\right|_T  = \left.
\frac{\pa U}{\pa \Om}\right|_T  =m r^2 \Om\, , \eea
which is the angular momentum, as expected since it is the
conjugate variable of the angular velocity $\Om$.

The above results may suggest that the condition
(\ref{thermalcondition}) can be then interpreted as a condition
for  an adiabatic physical system. $\frac{2 J_{2}}{\hbar}$ might
be viewed as an analogue of the Kolmogorov--Sinai entropy for
chaotic dynamical systems.

Finally, I note that a reparametrization-invariant time technique
in a specific model \cite{Elze:2002eg} also may lead to a quantum
dynamics emerging from a deterministic classical evolution.

\section{Conclusions}

In this report I have  reviewed some aspects of the algebraic
structure of QFT related with the doubling of the degrees of
freedom of the system under study. I have shown how such a
doubling is related to the characterizing feature of QFT
consisting in the existence of infinitely many unitarily
inequivalent representations of the canonical (anti-)commutation
relations and how this is described by the $q$-deformed Hopf
algebra. I have considered several examples of systems and shown
the analogies, or links, among them arising from the common
algebraic structure of the $q$-deformed Hopf algebra.

I have considered the Wigner function and the density matrix
formalism and shown that it requires the doubling of the degrees
of freedom, which thus appears to be a basic formal feature also
in Quantum Mechanics. In this connection I have considered the
two-slit experiment and shown that quantum interference effects
disappear in the limit of coincidence of the doubled variable
$x_{\pm}$. Then I have  shown how in QFT it is the $q$-deformed
coproduct which is relevant and how Bogoliubov transformations are
constructed in terms of it.  I have considered quantum dissipation
by studying the damped harmonic oscillator and the quantum
Brownian motion and commented on how the arrow of time emerges
from the intrinsic thermodynamic nature of dissipation. The vacumm
structure is the one of the generalized coherent states. The
connection (links) with the two-mode squeezed states and the
noncommutative geometry in the plane emerges in a natural way in
the discussion of these systems. In view of the similarity of some
features of the coherent states with those of the fractals, it is
an interesting question to ask whether fractal properties enter
the QFT structure. A study on this point is in progress.

The relation with thermal field theory, in the thermo field
dynamics formalism, reveals one further formal analogy with the
systems mentioned above. In such a contest entropy appears to be a
measure of the degree of entanglement between the system and the
thermal bath in which it is embedded. This also relates with the
connection between the doubled variables and quantum noise
effects.

For brevity, here I  have not considered the doubling of the
degrees of freedom in expanding geometry problems (inflationary
models) and in  the quantization of the matter field in a curved
background. For this I refer to the papers
\cite{Alfinito:2000bv,Martellini:1978sm,Iorio:2004bt}.

Finally, I have discussed how  't Hooft proposal, according to
which the loss of information due to dissipation in a classical
deterministic system manifests itself in the quantum features of
the system, finds a possible description in the formal frame
common to the systems mentioned above. In particular, I have shown
that the quantum spectrum of the harmonic oscillator can be
obtained from the dissipative character of the underlying
deterministic system. In recent years, the problem of quantization
of a classical theory has attracted much attention in gravitation
theories and in non-hamiltonian dissipative system theories, also
in relation with noncommutative space-time structures involving
deformation theory (see for example \cite{Aschieri:2005yw}). By
taking advantage of the fact that the manifold of the QFT
unitarily inequivalent representations is a K\"ahlerian manifold,
I have shown that classical trajectories in such a manifold, which
may exhibit chaotic behavior under some conditions, describe
(phase) transitions among the inequivalent representations. The
space of the QFT representations appears thus covered by a {\it
classical blanket}.

\bigskip
\bigskip
\bigskip

\noindent{\bf Acknowledgements}

\bigskip

I thank the MIUR, INFN and the ESF Program COSLAB for  partial
financial support. I am grateful to the organizers of the COSLAB
Dresden Workshop in July 2005 and in particular to Ralf
Schuetzhold for giving to me the opportunity to report in that
occasion on the matter here presented.

\bibliography{apssamp}

\printindex
\end{document}